# Worldwide Scientific Landscape on Fires in Photovoltaic

Esther Salmerón-Manzano [1], David Muñoz-Rodríguez [2], Alberto-Jesus Perea-Moreno [2,*], Quetzalcoatl Hernandez-Escobedo [3], Francisco Manzano-Agugliaro [4]

[1] Department of Law, University of Almeria, ceiA3, 04120 Almeria, Spain; esther.salmeron@ual.es

[2] Departamento de Física Aplicada, Radiología y Medicina Física, Universidad de Córdoba, Campus de Rabanales, 14071 Córdoba, Spain; qe2murod@uco.es ; aperea@uco.es

[3] Escuela Nacional de Estudios Superiores Unidad Juriquilla, Universidad Nacional Autónoma de Mexico, Blv. Juriquilla 3001, Queretaro 76230, Mexico. qhernandez@unam.mx

[4] Department of Engineering, University of Almeria, ceiA3, 04120 Almeria, Spain; fmanzano@ual.es

* Corresponding author: aperea@uco.es

## Abstract

The rapid growth of photovoltaic (PV) technology in recent years called for a comprehensive assessment of the global scientific landscape on fires associated with PV energy installations. This study examines the scientific literature indexed in Scopus from 1983 to 2023. It reveals a striking increase in output since 2011, with nearly one hundred publications in the most recent year under review. This growth of interest has occurred in parallel with the global expansion of photovoltaics. The majority of studies in this field are classified as engineering, with 34% of publications in this area. The USA leads the way with over 160 publications, followed by China with 125. Two institutions in the USA are particularly prominent in this field: Sandia National Laboratories in New Mexico with 22 publications, and the National Renewable Energy Laboratory in Colorado with 16 publications. The second institution is the University of Science and Technology of China, which has published 17 articles on the subject. A close examination of the evolution of keywords reveals a remarkable transformation in the scientific landscape over the past 10 years, from 2013 to 2023. The evolution of keywords suggests a maturation in the understanding of fire risks associated with photovoltaic energy. A total of seven scientific communities have been identified in which these works are grouped according to their keywords. These include Fire and Energy Storage, PV faults, Fire resistance, Fire hazard, Fire detectors, Deep learning, and Fire safety. It has been found that fires caused by PV installations are not listed as a cause of fire starts. This should be taken into account when conducting preventive analyses of this potential danger, particularly in light of the possible development of agrivoltaics, where facilities will be mainly located in the natural environment.

## 1. Introduction

Energy production is paramount for human development, influencing economic growth, technological innovation, healthcare, education, and overall quality of life (Baños et al., 2011). It serves as a catalyst for economic activities, job creation, and scientific progress, powering industries, businesses, and essential services like healthcare and education. Access to energy enhances agricultural practices, ensures food security, and supports the infrastructure that underpins modern societies (Juaidi et al., 2016). Additionally, energy is crucial for communication networks and connectivity (Arrubla-Hoyos et al., 2022) playing a pivotal role in maintaining global collaborations and information exchange. Striking a balance between meeting energy needs and promoting environmental sustainability is a central challenge for sustaining human progress into the future (Montoya et al., 2017).

Renewable energies play a crucial role in the global energy landscape, contributing significantly to sustainability and the principles of a circular economy (Manzano-Agugliaro et al., 2013). Firstly, renewable energy sources such as solar, wind, hydro, and geothermal power are essential in reducing dependence on finite fossil fuels, curbing greenhouse gas emissions, and mitigating climate change (Hernandez-Escobedo et al., 2011). According to the International Energy Agency (IEA), renewable energy capacity will allow to achieve a 50% reduction of global $CO_2$ emissions in the next years, representing around 90% of the total global power capacity increase. The sustainability aspect is evident in their ability to harness naturally replenishing resources, minimizing environmental impact and promoting long-term resilience (Bilgili et al., 2015). Moreover, the adoption of renewables aligns with the principles of a circular economy by emphasizing resource efficiency and minimizing waste (Perea-Moreno et al., 2017). The renewable energy sector also drives economic growth by creating jobs, fostering innovation, and attracting investments. The number of direct energy-related jobs in the power, heat, transport, and desalination sectors is projected to experience a significant rise, climbing from approximately 57 million in 2020 to nearly 134 million by the year 2050 (Ram et al., 2022), highlighting its role in job creation and economic development. Embracing renewables not only addresses energy challenges but also contributes significantly to building a sustainable and circular economic model.

Photovoltaic (PV) energy, a key component of renewable energy, holds significant importance in the global pursuit of sustainable power sources. Photovoltaic technology, commonly known as solar power, harnesses energy from the sun and converts it into electricity. Its importance lies in its remarkable environmental benefits and growing prominence in the global energy landscape. According to the World Bank, installed solar photovoltaic capacity has grown exponentially over the past decade, from 63669 GWh in 2011 to 1020439 GWh in 2021, a more than 16-fold increase, see figure 2, where the data of installed solar PV in that period are shown (The World Data Bank, 2024). The International Energy Agency (IEA, 2022) predicts that by 2027, Solar PV will have the world's largest installed power capacity, surpassing coal. According to their forecast, cumulative solar PV capacity is expected to almost triple, increasing by almost 1,500 GW over the period. This growth is projected to exceed that of natural gas in 2026 and coal in 2027. The most important aspect of photovoltaics is its inherent sustainability, as it generates electricity without the emission of greenhouse gases, thus contributing to the reduction of the carbon footprint and the fight against climate change. Solar energy is also characterised by decentralisation, allowing power generation at the point of consumption, which increases energy security and resilience (Albatayneh et al., 2022a). In addition, the declining costs of PV technology have made solar energy increasingly competitive, further driving its adoption and making it a cornerstone of the transition to a cleaner and more sustainable energy future (Albatayneh et al., 2022b).

Although photovoltaics (PV) has numerous environmental and economic benefits, a major drawback is their association with various types of fires (Ju et al., 2018). Internal issues are responsible for 50% of fires in photovoltaic systems located in roof (Ong et al., 2022). These issues arise from faults in the installation itself, such as faulty element installation, overheating of poorly ventilated panels or inverters, and electrical faults due to poor wiring or faulty cable insulation. External factors can also contribute to the causes of these fires, such as extreme environmental conditions such as prolonged droughts or nearby fires, physical impacts such as falling objects, hail, storms, or animals that damage solar panels or cables, and lack of maintenance. The increasing use of lithium-ion batteries in energy storage systems also poses a fire risk (Ghi et al., 2021). There is a lack of comprehensive data on fires caused by PV installations, which are usually classified as 'other' incidents. As a reference, a frequency analysis shows 0.289 fires per MW installed, or 28.9 fires per GW installed (Ong et al., 2022).

Potential fire hazards in PV systems are a critical concern that requires thorough analysis and mitigation strategies (Juarez-Lopez and Franco, 2023). Extinguishing a fire in a photovoltaic electrical installation is a challenging task. This is due to the unique characteristic of photovoltaic modules, which continue to produce energy despite the presence of fire. It is important to note that electricity continues to flow through the installation, even during a fire. For instance, the cables of a photovoltaic installation can carry up to 1500V of direct current. In other words, water alone cannot extinguish a fire as it poses an added danger of electrocution. Additionally, it is important to consider that photovoltaic systems are often situated in remote or hard-to-reach locations, such as rooftops.

One of the first studies to look at the large scale viability of PV roofs was Boumans et al. (1994), who analysed this issue for the Netherlands. They studied two PV roof integration systems and the results showed that the PV quality of these systems could be guaranteed for more than 5 years. As for the evaluation of fire resistance, tests were carried out according to the NEN 6063 standard and the ignition area did not increase. It was not until 25 years later that a second study was published, dealing with the integration of photovoltaic roofs and analysing fire resistance as one of their possible limitations (Cross, 2018). Recent studies have made it possible to verify the fire performance of some PV modules in combination with different types of roofing materials and have helped to identify some weaknesses in the European and Italian test protocols for fire performance qualification, mainly due to the specific design and outdoor installation characteristics of PV modules, especially in aspects such as module tilt, module power or ignition flame duration (Cancelliere et al.,2021).

When considering remote locations, it is important to consider the development of agrivoltaics. This involves the compatibility of energy production from photovoltaic modules with agricultural production. The first experiments on agrivoltaics were carried out in 2011 in Montpelier (France) by Dupraz et al. (2011) who found increases in land productivity of up to 60-70%. It is important to note that 90% of the area covered by the PV system is considered to be cultivated. Only the proportion of the area occupied by the PV pillars is deducted. This is a promising future for photovoltaics. Especially in countries where farmland is protected because of the direct relationship between food production and arable land. The average amount of arable land available per person worldwide is 0.19 ha, ranging from 0.05 to 2 ha per person depending on the country concerned (Salmeron-Manzano & Manzano-Agugliaro, 2023). The concept of agrivoltaics is a more sustainable approach than the traditional method of confining a photovoltaic installation to the ground, as it is achieved through the study of maximum energy production and the avoidance of shading of the panels depending on their inclination (Castellano et al., 2015).

Understanding and addressing these fire-related challenges is crucial to ensure the safety and reliability of PV systems (Szultka et al., 2023). Research and analysis of global trends in PV-related fires are essential to develop safety standards, guidelines and technologies that can mitigate these risks and improve the overall safety of solar energy installations. Balancing the benefits of clean energy with the need for robust safety measures is key to promoting the widespread adoption of PV. This is crucial even in agricultural areas, as previously mentioned.

Aram et al. (2021) conducted a review of studies and expert reports on fire safety in PV systems. The aim was to identify actual fires in PV panel systems and detect possible errors in the PV panel system elements that could increase the pre-existing fire risk. The aim of our study is to analyse the scientific landscape on fire and photovoltaics to identify global research trends in terms of number of publications, areas of expertise, affiliations and countries. In addition, the study aims to identify the scientific communities involved in these research areas through analysis of keywords used in the literature. These clusters will be identified through computational analyses of various data points, such as citation networks, or shared keywords in research publications. By mapping these connections, researchers can uncover the structure of the scientific field, revealing hubs of collaboration and areas of specialized research interest.

## 2. Methodology

The Scopus scientific database plays a pivotal role in supporting research endeavours by offering comprehensive, multidisciplinary, and reliable information, coupled with advanced tools for analysis and evaluation (Hernandez-Escobedo et al., 2018). Its global reach and integration capabilities make it an indispensable resource for the research community. Researchers can use Scopus to monitor trends in their fields, identify emerging topics, and stay abreast of the latest developments (Salmerón-Manzano, 2021). This assists in shaping research agendas and contributing to the advancement of knowledge. The database offers sophisticated search functionalities, allowing researchers to tailor their queries to specific criteria. This helps in refining searches and obtaining relevant results efficiently. In this research the search query was: (TITLE-ABS-KEY (fire*) AND (TITLE-ABS-KEY (photovoltaic*) OR TITLE-ABS-KEY (PV))). The obtained data from Scopus has been exported in a suitable format to be analyzed, see Figure 1.

The problem of community detection arises from a common feature inherent to all complex systems, which is the presence of patterns of nodes that are more densely connected to each other than to the rest of the nodes in the network (Guerrero et al., 2017). These densely connected nodes are called communities, and they are expected to share certain properties that allow the detection of new features or functional relationships of the network (Guerrero et al., 2018). Finding these patterns or community structures is known as the problem of community detection. Finding the optimal community structure that best represents the characteristics of the network has become a scientific challenge. To this end, a variety of algorithms and objective functions have been proposed to solve the problem, with evolutionary algorithms and the modularity index standing out as the main solutions accepted by the scientific community (Guerrero et al., 2019).

The Vosviewer software tool applies an algorithm designed for modularity-based community detection in high-volume networks. The concept of modularity functions was originally introduced by Newman and Girvan, with Newman himself proposing their use in community detection through the optimisation of a modularity function (Waltman and Van Eck, 2013). There are different adaptations of the modularity-based approach to community detection that address targeted or weighted networks and provide a resolution parameter (Newman, 2004). This parameter allows to adjust the level of granularity at which communities are identified and addresses the so-called resolution limit problem (Leicht and Newman, 2008).

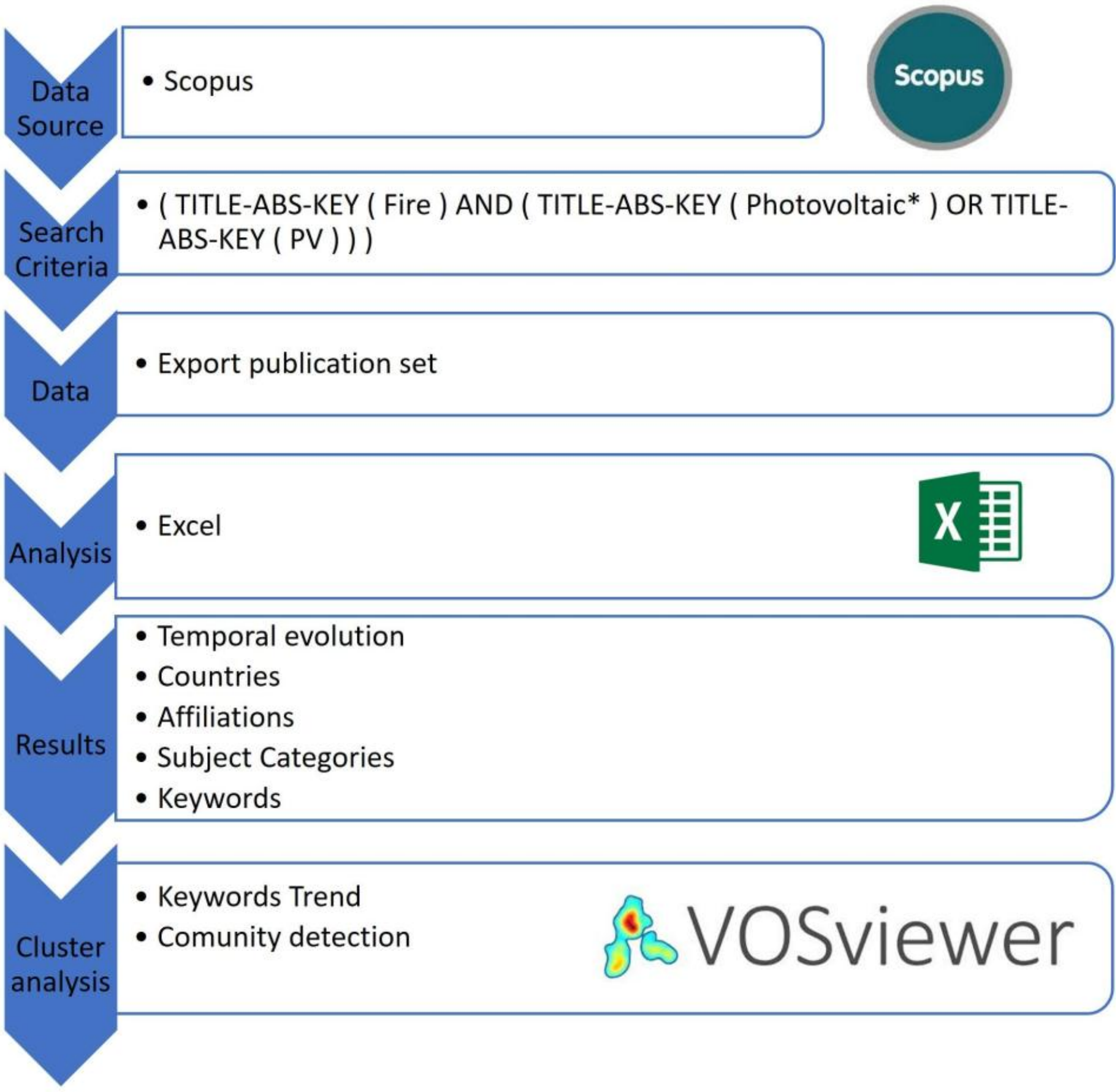

**Figure 1.** Methodology chart.

## 3. Results

The search has resulted in more than 700 published works in the last 40 years, from 1983 to 2022. The evolution trend in the research study, as reflected in the obtained data, reveals a substantial growth in research output over the years. The early years (1981-1990) show a relatively low number of studies, with some fluctuations. The field was likely in its nascent stage, and research interest was gradually building.

Figure 2 shows the evolution of these publications from 1990. As can be seen in the graph, the time trend of publications on this topic is quite similar to the trend of photovoltaic power installations, also worldwide. After, years 1990-2003, a noticeable increase in the number of studies is observed from the early 1990s to the early 2000s. This could be indicative of growing recognition and interest in biometric studies during this period. The period 2004-2010, shows some fluctuations in the number of studies during this period, suggesting possible shifts in research focus or periodic variations in interest.

A substantial increase in scientific output is observed from 2011 to 2014. This period marks a significant acceleration in the research topic, possibly driven by technological advances, increased applications or emerging trends in the field of photovoltaics and the resulting fire risk. From 2015 onwards, there is a steady upward trend in the number of studies. The field of study appears to have matured, attracting sustained attention and contributing to a growing body of knowledge. Between 2018 and 2022, the highest number of studies is recorded, indicating a peak in research activity, this could be attributed to the growing of global photovoltaic installations. The data for 2022, not shown in the figure 2, shows a marked increase to 95 studies, this is indicative of a continued high level of interest and may reflect the potential fire problem associated with the installation of photovoltaic systems.

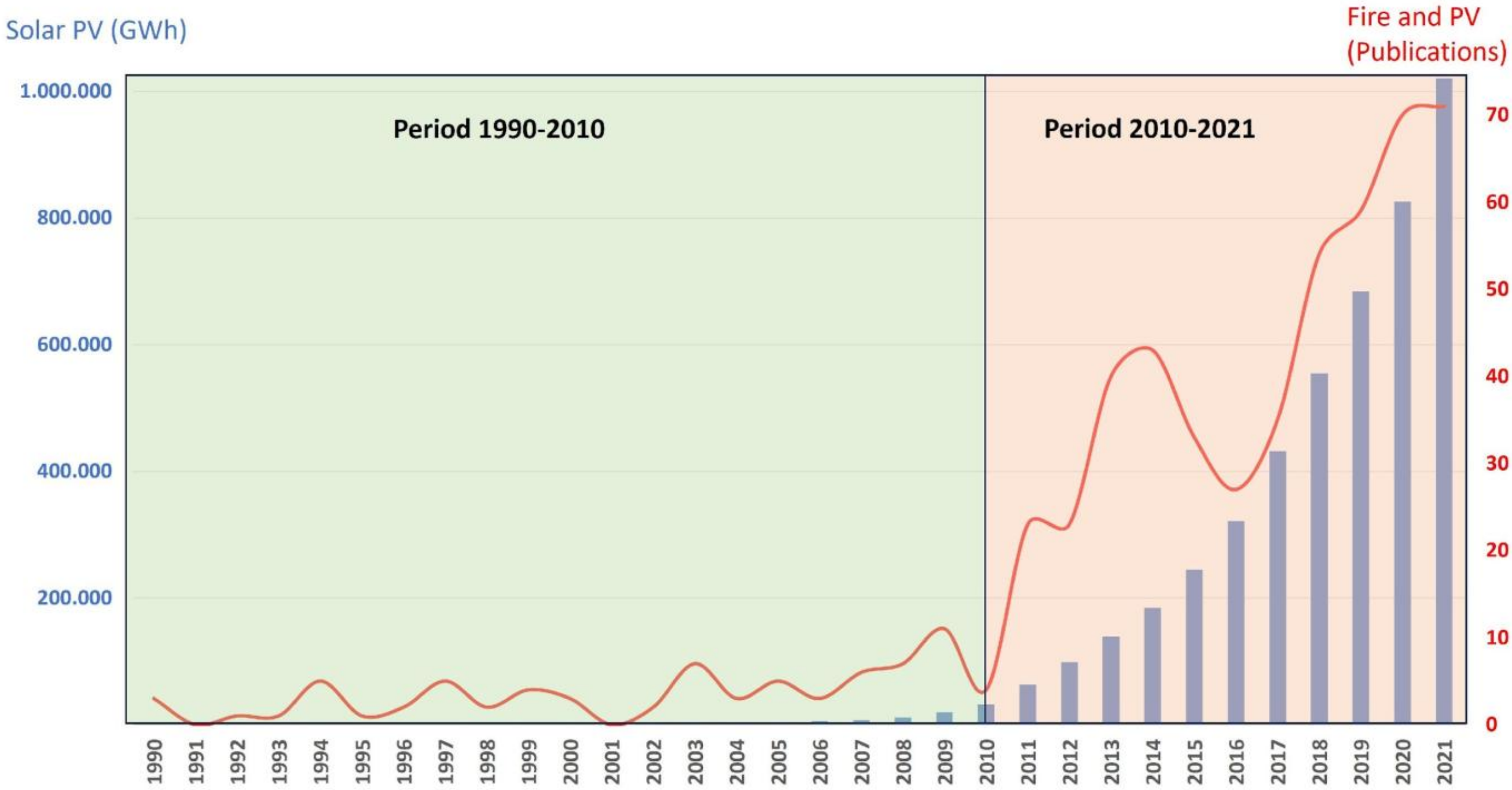

**Figure 2.** Evolution Trend of Installed Solar PV and publications related to Fires and PV.

Table 1 shows the main keywords appearing in these publications, and Figure 3 shows a word cloud to highlight their relative importance. The set of keywords with their respective occurrences in the scientific articles reflects the various aspects covered in the bibliometric analysis related to fire and photovoltaics. Keywords such as "Photovoltaic cells", "Photovoltaic systems" and "Solar power generation" indicate a strong focus on the technological and generation aspects of photovoltaics. There is a notable emphasis on fire-related terms such as "Fires", "Fire hazards" and "Fault detection", suggesting a substantial interest in the study and prevention of fires in PV systems. Terms such as "Fire protection", "Safety" and "Electrical grounding" highlight the concern for safety measures and protection strategies in PV installations. The inclusion of "Reliability", "Risk assessment" and "Fault diagnosis" points to an interest in assessing the reliability and potential risks associated with PV systems. Keywords such as "Optimisation", "Energy efficiency" and "Maximum power point tracking" indicate an interest in optimising the performance and energy efficiency of PV systems. Terms such as "MATLAB", "Signal Processing", "Machine Learning" and "Artificial Intelligence" suggest the use of advanced technological tools and methods for analysis and detection. Keywords like "Renewable Energies," "Environmental Impact," and "Sustainable Development" reflect an awareness of the environmental impact and sustainability considerations in the context of photovoltaic energy and fires. Inclusion of terms like "Deep Learning," "Neural Networks," and "Particle Swarm Optimization (PSO)" suggests an exploration of advanced technologies for analysis and optimization.

**Table 1.** Main Keywords.

| Keyword | N |
|---|---|
| Photovoltaic Cells | 260 |
| Photovoltaic Systems | 127 |
| Solar Power Generation | 114 |
| Fault Detection | 114 |
| Fires | 103 |
| Solar Energy | 86 |
| Photovoltaic | 68 |
| Photovoltaics | 67 |
| Fire Hazards | 59 |
| Solar Cells | 54 |
| Fire Protection | 54 |
| Photovoltaic System | 51 |
| Photovoltaic Effects | 47 |
| Photovoltaic Modules | 44 |
| Photovoltaic Arrays | 44 |
| PV System | 35 |
| Solar Panels | 33 |
| Smoke | 32 |
| Maximum Power Point Trackers | 31 |
| Arc Faults | 30 |

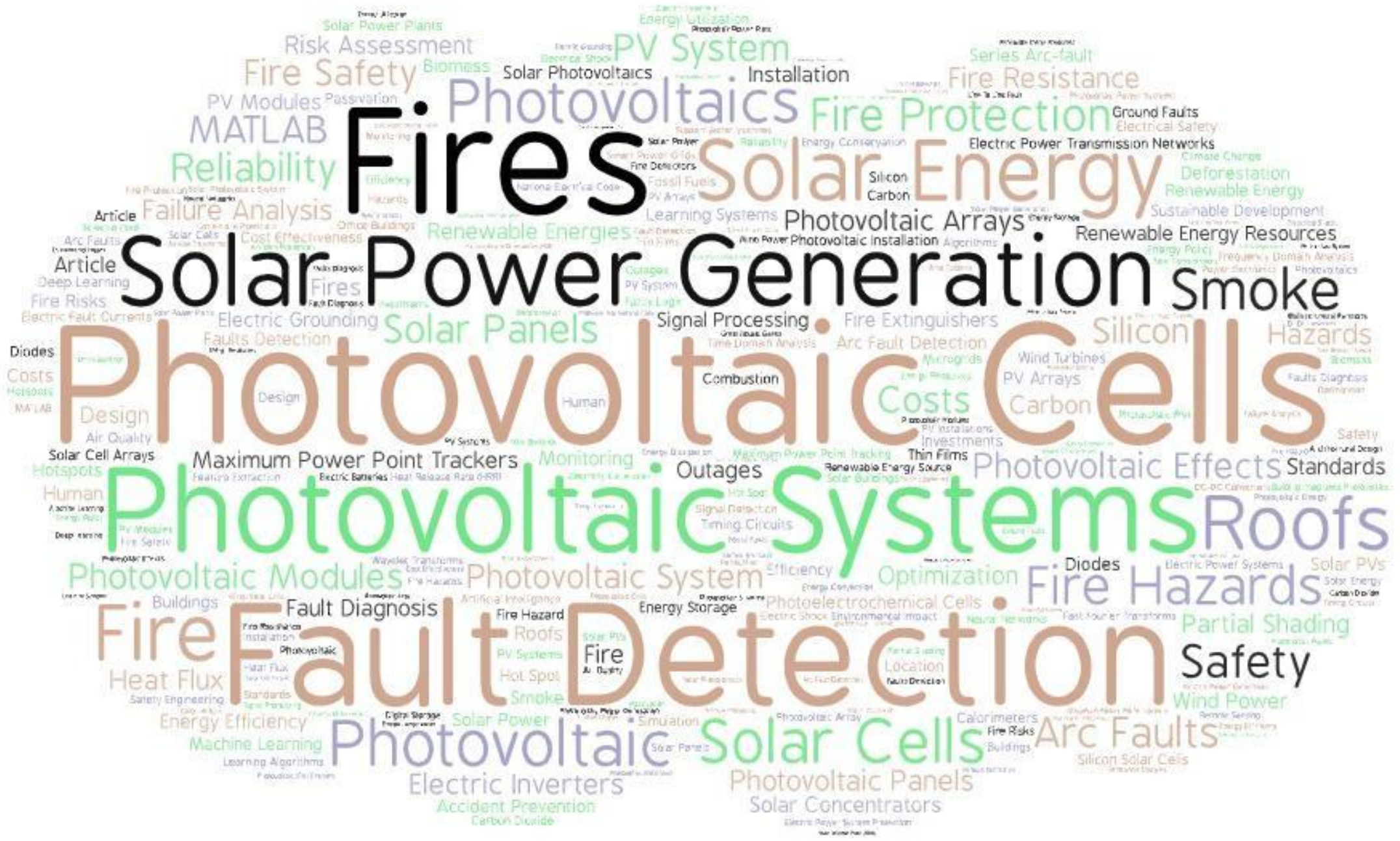

**Figure 3.** Cloud words of Fire vs Photovoltaic.

An analysis of the results by scientific subject category shows in Figure 4 that 34% of the publications fall into the scientific category of Engineering, followed by Energy with half of these publications, i.e. 17%. Material Science and Computer Science categories with 10% each.

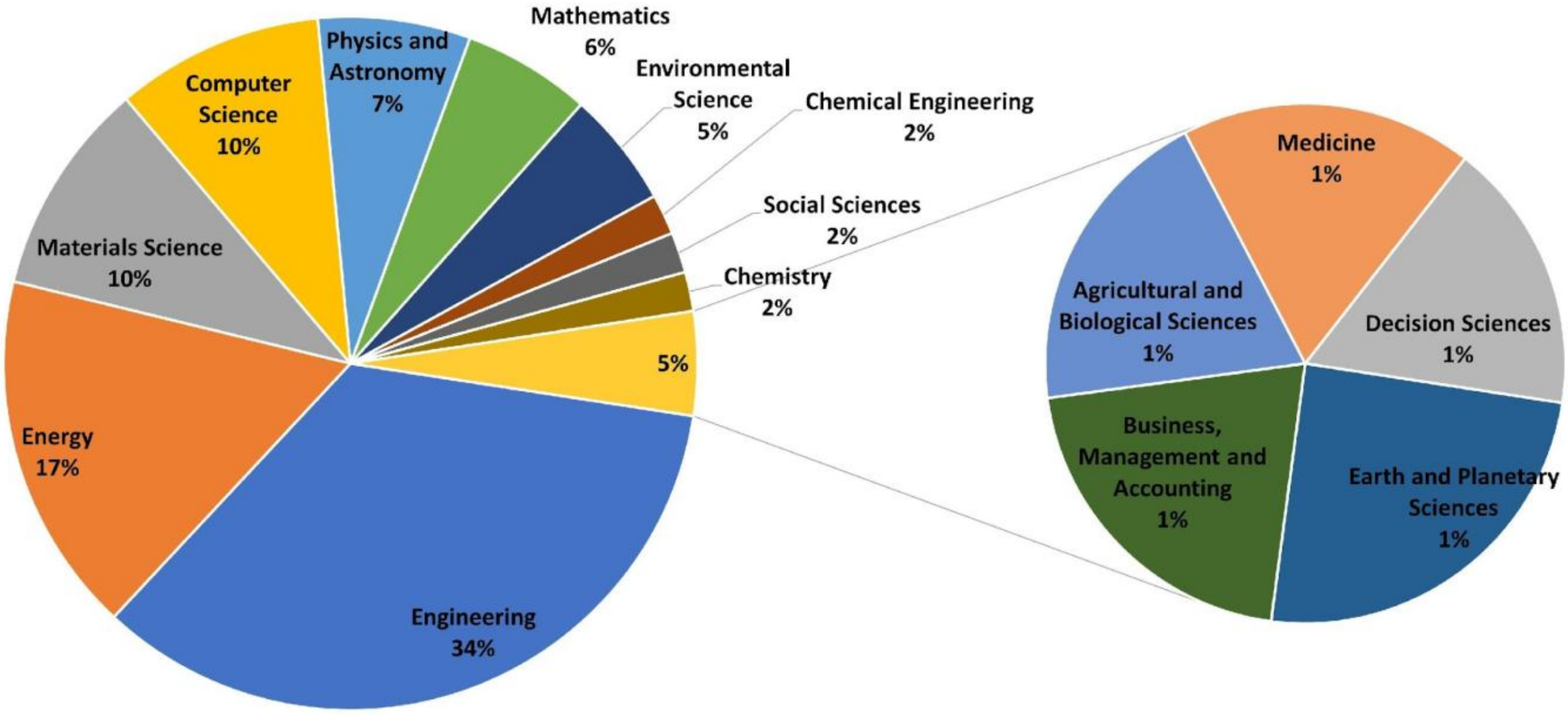

**Figure 4.** Distribution by scientific category of the publications in fire and PV.

### 3.1. Countries and their main topics

The country data from the bibliometric study on fires related to photovoltaic energy reveals interesting patterns and variations in research output across different nations, see figure 5. The distribution of research output across countries highlights the global nature of interest in the safety aspects of photovoltaic energy, with certain nations taking a lead in contributing to the scholarly discourse on this topic. The variations in output may be influenced by factors such as technological adoption, policy emphasis, and national research priorities. The United States (<150) leads in research output, which is consistent with its prominent role in scientific research and technological innovation. China (≈ 125) follows closely behind, indicating the country's growing emphasis on research in the field of fires related to photovoltaic energy. This aligns with China's significant investments in renewable energy technologies. India's representation (≈60) in the study signifies a noteworthy level of research activity in the context of fires associated with photovoltaic energy, reflecting the country's increasing focus on renewable energy sources. Germany (30), known for its strong commitment to renewable energy, demonstrates a substantial but relatively lower research output compared to larger countries like the United States and China. Italy, United Kingdom, and Japan (<30), these countries show similar research outputs, suggesting a shared interest and engagement in understanding and mitigating fire risks in the context of photovoltaic energy. Canada, South Korea, and Taiwan (≈20), these nations exhibit moderate research output, reflecting a certain level of engagement and awareness regarding the intersection of fires and photovoltaic energy. France, Spain, and Iran (<20), these countries contribute to the research landscape with a notable but somewhat lower number of publications. The numbers may indicate varying levels of emphasis on this specific aspect of photovoltaic energy. Australia and Malaysia (<15), these countries show a similar level of research output, suggesting a common interest in studying fires related to photovoltaic energy. Belgium and Denmark (10), while these countries have a lower research output, it indicates their involvement in understanding and addressing fire issues in the context of photovoltaic energy.

The keywords chosen by each country provide insights into their specific research priorities within this interdisciplinary field, see Table 2. The data reveals a diverse research landscape across countries, with varying emphasis on technological aspects, safety considerations, and architectural concepts related to fires and photovoltaics.

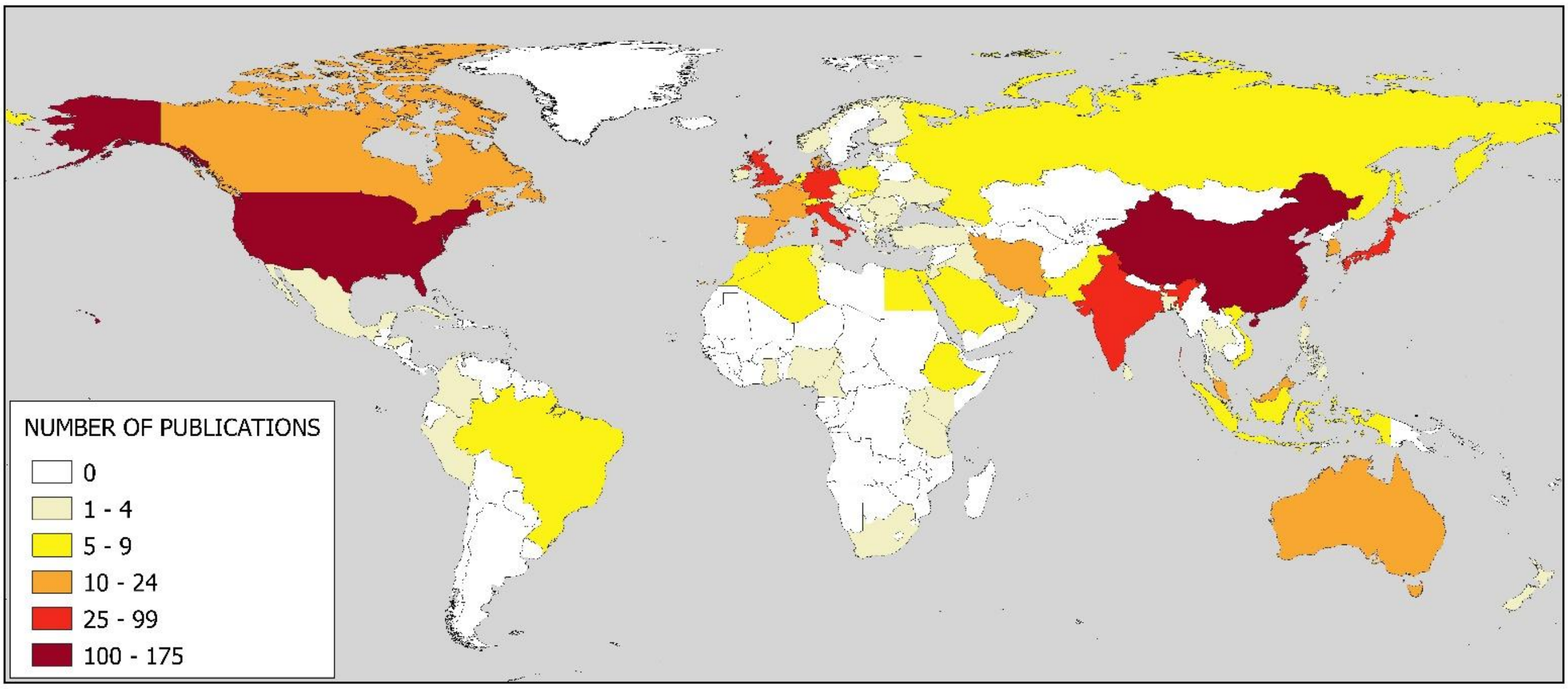

**Figure 5.** Geographical distribution by country of the worldwide publications on fires and photovoltaic.

**Table 2.** Countries and their main topics.

| COUNTRY | N | Keyword-1 | Keyword-2 | Keyword-3 | Keyword-4 |
|---|---|---|---|---|---|
| United States | 161 | Photovoltaic Cells | Fires | Solar Power Generation | Photovoltaic |
| China | 125 | Photovoltaic Cells | Fault Detection | Photovoltaic Systems | Fires |
| India | 63 | Photovoltaic Systems | Fault Detection | Solar Power Generation | Fire Hazards |
| Germany | 30 | Photovoltaic Cells | Fires | Photovoltaic Systems | Fault Detection |
| Italy | 29 | Photovoltaic Cells | Fires | Photovoltaic Systems | Risk Assessment |
| United Kingdom | 29 | Photovoltaic Cells | Fires | Roofs | Solar Cells |
| Japan | 28 | Photovoltaic Cells | Solar Cells | Solar Energy | Solar Power Generation |
| Canada | 22 | Fires | Smoke | Fire Safety | Photovoltaic Cells |
| South Korea | 20 | Photovoltaic Cells | Renewable Energies | Accidents | Photovoltaic |
| Taiwan | 19 | Photovoltaic | Fault Detection | Fire Hazards | Photovoltaic |
| France | 17 | Photovoltaic Cells | Buildings | Fire Hazards | Fires |
| Spain | 16 | Photovoltaic Cells | Fires | BIPV | Building Integrated |
| Iran | 15 | Fault Detection | Photovoltaic Cells | Photovoltaic Systems | Photovoltaic Arrays |
| Australia | 14 | Australia | Photovoltaic Systems | Photovoltaic Cells | Solar Energy |
| Malaysia | 14 | Photovoltaic Systems | Fault Detection | Photovoltaics | Solar Power Generation |
| Belgium | 10 | Photovoltaic Cells | Heat Flux | Roofs | Architectural Concepts |

Figure 6 shows the scientific subject categories in which the three main countries publish. The USA uses 42% of these publications in Engineering, while China uses 32% and India 36%. In terms of the Energy category, China devotes the most effort with 20% of its publications, while the USA devotes 16% and India 15%. In terms of publication distribution, India leads in Computer Science with almost 17%, followed by China with 9% and the USA with 8%. In Environmental Sciences, China has the highest percentage with 6%, while the USA and India have 5% and 3%, respectively.

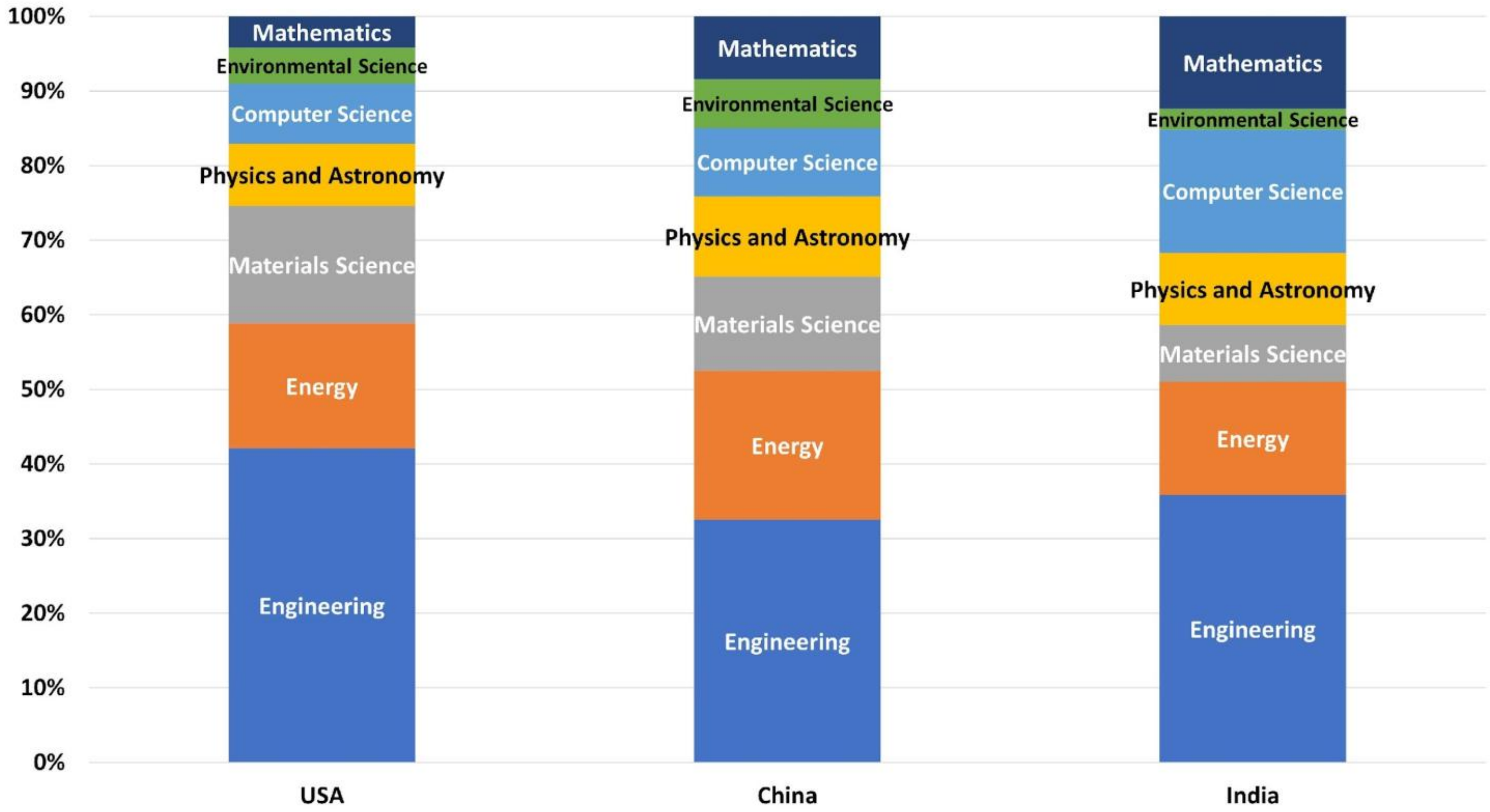

**Figure 6.** Distribution by subject category for the top 3 countries.

### 3.2. Affiliations and their main topics

The affiliation data from the bibliometric study on fires related to photovoltaic energy highlights the institutions that have actively contributed to research in this area. The affiliation data highlights a mix of national laboratories, universities, and governmental organizations actively involved in the research on fires in the context of photovoltaic energy. The collaboration between academic institutions, research laboratories, and governmental bodies underscores the interdisciplinary nature of this research area.

Sandia National Laboratories (22 publications) is a prominent contributor, indicating a significant role in research related to fires and photovoltaic energy. This could be due to its focus on energy research and national security. The University of Science and Technology of China (17 publications) is a leading academic institution contributing substantially to the research on fires in the context of photovoltaic energy. This reflects China's commitment to advancing research in renewable energy. The National Renewable Energy Laboratory's high research output underscores its central role in studying renewable energy sources, particularly in understanding and addressing fire-related issues in photovoltaic systems (16 publications). The presence of Université de Sherbrooke (11 publications) among the top affiliations suggests its active involvement in research related to fires and photovoltaic energy. This may be indicative of the institution's focus on renewable energy studies. The Ministry of Education of the People's Republic of China (10 publications) is an interesting inclusion, suggesting the involvement of governmental bodies in research initiatives related to the safety aspects of photovoltaic energy. Southeast University's publications (9 publications) indicate their contribution to the scholarly discourse on fires in the context of photovoltaic energy, reflecting its research activities in the renewable energy domain. Texas A&M University's presence (9 publications) highlights its research contributions to understanding and mitigating fire risks associated with photovoltaic systems, showcasing the importance of academic institutions in this field. The National Institute of Technology Tiruchirappalli's inclusion suggests an active engagement in research on fires related to photovoltaic energy, reflecting the global nature of interest in this topic (8 publications). Shanghai University and Nanjing Tech University (8 and 6 publications), these Chinese universities contribute significantly to the research output, emphasizing the role of academic institutions in advancing knowledge in this specific area. The University of Edinburgh, Brookhaven National Laboratory, Technical University of Denmark (7, 6, and 6 publications), and these institutions from different countries demonstrate a collective effort in researching fires and photovoltaic energy, showcasing the global collaboration and diverse expertise involved in this field.

Photovoltaic (PV) systems have witnessed widespread adoption globally, significantly contributing to renewable energy generation. However, the increasing integration of these systems has drawn attention to the critical issue of fire incidents in PV installations. This compilation focuses on the most cited research articles across various countries, shedding light on diverse aspects of fire-related challenges in photovoltaic arrays. These publications delve into catastrophic failures, fault detection methodologies, environmental impacts, and cutting-edge technologies aimed at ensuring the reliability, safety, and efficiency of PV systems. The urgency of addressing fire risks in PV installations is underscored by the prominence of these articles, recognized as influential contributions in the field.

The most cited publication from the United States explores catastrophic failures in PV arrays, including ground faults, line-to-line faults, and arc faults. Notably, recent fire incidents in Bakersfield, CA, USA, in 2009, and Mount Holly, NC, USA, in 2011 underscore the necessity for improved fault detection and mitigation techniques. This study aims to comprehend the impact of faults on PV array operation, identify current limitations in detection and mitigation methods, and provide a survey of advanced fault detection technologies and commercially available products (Alam and Flicker, 2015).

In the case of China, Wu et al. (2013) address the impact of partial shading on PV systems, a factor that can reduce output power and lead to issues like hot spots, potentially causing cell damage or fires. The study explores the output characteristics of PV modules under partial shading and proposes a method using slope and short-circuit current for hot spot detection. Additionally, the authors introduce a fault diagnosis and optimal control strategy based on fuzzy control to extend the service life of PV modules. Simulation and experimental results validate the effectiveness of the proposed approach.

The most-cited publication from India brings attention to the insufficient focus on fault detection in PV systems despite the growing global PV capacity. The paper categorizes advanced fault detection approaches, considering factors such as types of faults detected, detection time, sensor requirements, procedural complexity, detection variables, and the level of protection achieved. This research emphasizes the need for improved fault detection possibilities in PV systems and serves as a valuable reference for researchers in this field (Pillai et al., 2019).

The highest-cited paper from Germany discusses the risk of arc faults in photovoltaic systems due to high DC voltages and aging. It proposes sensor devices for detection and highlights challenges in identifying series arc faults. Analysis suggests the necessity of sensors and methods to detect higher spectral components of arc effects (Strobl & Meckler, 2010).

The publication with the most citations from Italy underscores the importance of detecting and diagnosing faults in PV systems for efficiency, energy yield, and overall safety. The study reviews various fault detection and diagnosis (FDD) methods, focusing on PV arrays, and highlights the challenges, advantages, and limitations of these methods in terms of feasibility, complexity, cost-effectiveness, and scalability for large-scale integration. Recommendations for future research directions are also provided (Mellit et al., 2018).

In the United Kingdom, the most frequently cited work focuses on addressing the issue of hot spots in outlier solar cells within PV modules, which can lead to accelerated aging and potential failure, including fire risks. The proposed solution is an improved bypass circuit that prevents hot spots without the need for preliminary detection, offering a simpler and more efficient alternative to standard bypass diodes. The circuit self-activates under normal diode conditions, eliminating the need for control logic or an external power supply (Guerriero et al., 2019).

The leading citation in Japan pertains to investigating solder joint failures in traditional PV modules using crystalline silicon solar cells. Two failure modes, Ag or Cu leaching into solder and long-term solder joint fatigue, can lead to cracks and potential DC arcing discharge. The study emphasizes the importance of safeguarding against solder joint failures to ensure the reliability and safety of PV module systems (Itoh et al., 2014).

The most cited research from Canada is centered on the environmental impact of prefabricated temporary housing with renewable energy systems, specifically solar PV systems. Utilizing a life cycle assessment (LCA) model, the study proposes carbon reduction measures, showcasing significant reductions in material embodied, assembly, and operational emissions. This research contributes to promoting sustainable development in prefabricated temporary housing with an emphasis on carbon reduction (Dong et al., 2018).

In the case of South Korea, Cho et al. (2009) investigate screen-printed silver thick-film metallization for crystalline silicon solar cell contacts. The study explores Ag crystallite formation and contact resistance under different oxygen partial pressures during firing treatments, revealing strong dependencies on the firing ambiance's oxygen levels.

The top-cited publication originating from Taiwan is about a novel fault diagnostic method for PV systems (PVS). The approach involves extracting optimal fault features, standardizing them under standard test conditions, and establishing a fault diagnostic model using an adaptive boosting algorithm. The method proves accurate and reliable in detecting various types of faults in PV modules, as demonstrated through numerical simulations and experimental results (Huang et al., 2019).

The publication with the highest number of citations in France proposes a graph-based semi-supervised learning model for fault detection in solar PV arrays. This model requires only a few labelled training data, addressing challenges associated with extensive data labelling. It not only detects faults but also identifies possible fault types, expediting system recovery. The proposed method proves effective in both simulation and experimental results (Zhao et al., 2019).

At the forefront of citations for Spain is a publication addressing a review of storage systems for Building-Integrated Photovoltaic (BIPV) and Building-Integrated Photovoltaic/Thermal (BIPVT) applications. The study explores batteries, Phase Change Materials (PCMs), and water tanks, emphasizing their environmental impact. It also evaluates alternative storage options, providing a critical overview of their suitability and environmental considerations (Lamnatou et al., 2020).

The publication from Iran that has received the highest number of citations proposes an intelligent fault diagnosis method for PV arrays, effectively detecting and classifying Line-Line (LL) faults using an ensemble learning model and Current-Voltage (I-V) characteristics. The method achieves high accuracy, outperforming individual machine learning algorithms with an average accuracy of 99% for fault detection and 99.5% for fault classification (Eskandari et al., 2020).

The most referenced paper originating from Australia presents a review of current techniques for detecting DC arc faults in PV systems. With the increasing integration of solar energy, improper installation and aging can elevate the risk of arc faults, posing fire hazards. The paper discusses methods and features for DC arc fault detection, emphasizing the importance of simulation in studying characteristics and fault diagnosis. Various DC arc fault models are also reviewed (Lu et al., 2018).

In Malaysia, the publication garnering the most citations is dedicated to proposing a novel design for Rooftop Mounted Photovoltaic (RMPV) systems, integrating real-time data processing for reliable supervision. Using multiblock Principal Component Analysis (PCA) and Kernel Density Estimation (KDE), they extract and monitor Dominant Transformed Components (TCs) to detect anomalies, including array faults, DC-side mismatches, grid-side issues, inverter anomalies, and sensor faults. The study utilizes historical data from various RMPV energy conversion systems during 2015–2017 (Bakdi et al., 2019).

Finally, the publication with the highest number of citations in Belgium introduces a machine learning-based approach for fault detection in PV systems. Leveraging satellite weather data and low-frequency inverter measurements, the method, powered by a recurrent neural network, identifies six fault types and estimates severity over the past 24 hours. With high accuracy (96.9%±1.3% with exact weather data and 86.4%±2.1% with satellite weather data), the model is sensitive to faults, generalizes across climates, and can detect unknown faults (Van Gompel et al., 2022).

## 4. Scientific landscape evolution

In figure 7, the evolution of keywords in this research topic is depicted. A notable shift can be observed, especially over the last 10 years, from 2013 to 2023. Between 2013 and 2015, research on fires related to photovoltaic energy was comprehensive, spanning a spectrum of critical themes. Initial attention was directed towards the integration of solar cells into buildings and structures, suggesting a focus on the practical applications of photovoltaic technology. Safety emerged as a primary concern, with investigations into electric shock risks and endeavours to enhance fire resistance within photovoltaic systems. The exploration extended to the realm of materials, particularly polymers, where researchers probed into their safety and durability. The testing of these materials, as well as the overall reliability of photovoltaic systems, was a prominent aspect of this period's research. The potential hazards of specific materials like cadmium, considerations related to heating, and the role of aluminium in photovoltaic systems rounded out the multifaceted examination. This timeframe reflects a holistic approach, encompassing safety, material science, and practical applications in the evolving landscape of photovoltaic energy.

During the years 2016 to 2017, research on fire incidents associated with photovoltaic energy honed in on specific and critical aspects. There was a noticeable pivot towards a more explicit examination of fire-related concerns, acknowledging an increasing awareness of the potential hazards linked to photovoltaic installations. The investigation extended to the impact of photovoltaic panels on roofs, underlining the importance of understanding their role in fire risk management. The inclusion of keywords like glass and architectural design suggests a nuanced exploration of how materials and design elements contribute to both the efficiency of photovoltaic systems and their potential fire risks. This period marked a transition towards a more detailed analysis of the relationship between fire hazards and the physical characteristics of photovoltaic installations, emphasizing the need for a holistic approach in ensuring both efficiency and safety within this evolving energy landscape.

In the years 2018 to 2019, research on fires linked to photovoltaic energy evolved towards a more technologically oriented investigation. The emphasis on fault detection and grid-connected systems suggests a growing awareness of the complexities associated with integrating photovoltaic installations into smart power grids. The inclusion of keywords like short circuits fault, fast Fourier transform, and DC side reflects a specialized focus on technical intricacies. Researchers likely delved into issues such as short circuit vulnerabilities, the application of Fourier transform for signal analysis, and specific challenges on the direct current (DC) side of photovoltaic systems. This period underscores a shift towards addressing the nuanced technical aspects and potential vulnerabilities associated with the integration of photovoltaic technology into interconnected and smart energy grids, indicating a maturation in the understanding of system dynamics and potential fire risks.

From 2020 to 2023, the landscape of research on fires associated with photovoltaic energy exhibits a pronounced shift towards advanced diagnostic and detection methodologies. The keywords such as fault diagnosis, computer-aided diagnosis, and detection methods suggest a keen interest in developing sophisticated approaches to identify and address potential issues in photovoltaic systems. The inclusion of Maximum Power Point Tracking (MPPT), hotspots, partial shading, and deep learning underscores a focus on optimizing energy output, addressing localized temperature anomalies, and employing advanced machine learning techniques for enhanced system monitoring and fault prediction. This period reflects a strategic move towards leveraging cutting-edge technologies to enhance the safety and efficiency of photovoltaic installations, indicating an awareness of the importance of proactive measures in mitigating fire risks and optimizing overall system performance.

The evolution of keywords suggests a maturation in the understanding of fire risks associated with photovoltaic energy, transitioning from broader concerns about materials and safety to more specific technical challenges and grid integration issues. This progression aligns with the broader development of PV technology and its integration into various sectors.

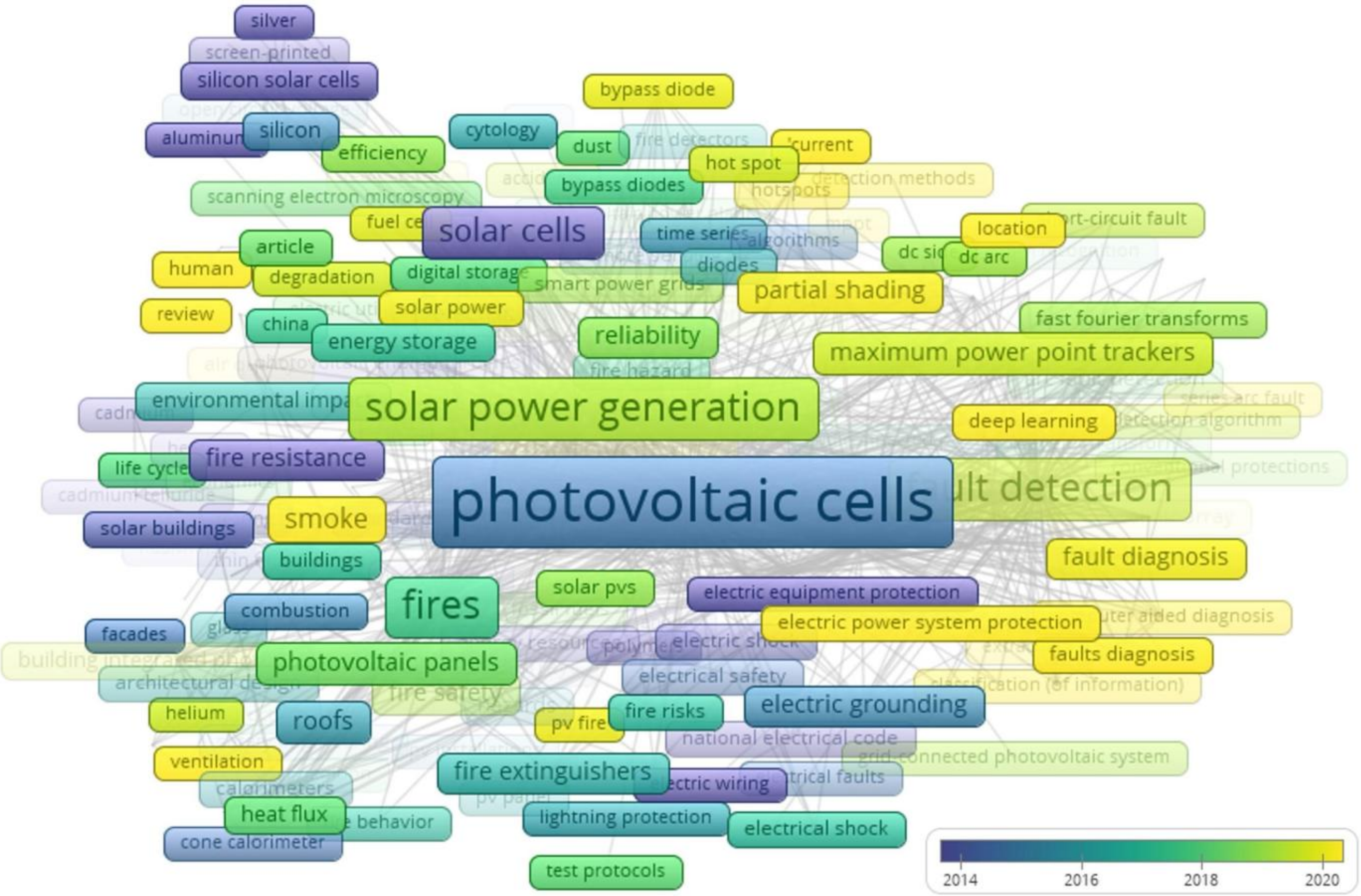

**Figure 7.** Scientific landscape evolution of keywords on Fire and Photovoltaic.

## 5. Discussion: Analysis of Scientific Communities or Clustering

Clustering allows researchers to identify patterns and trends within scientific communities. By grouping related publications together, it becomes easier to discern the direction in which research is progressing, helping to identify emerging themes and areas of interest. Figure 8 shows the seven clusters obtained for the search query and the analysis done. Table 3 shows the main keywords used by the communities detected in the topic of this study.

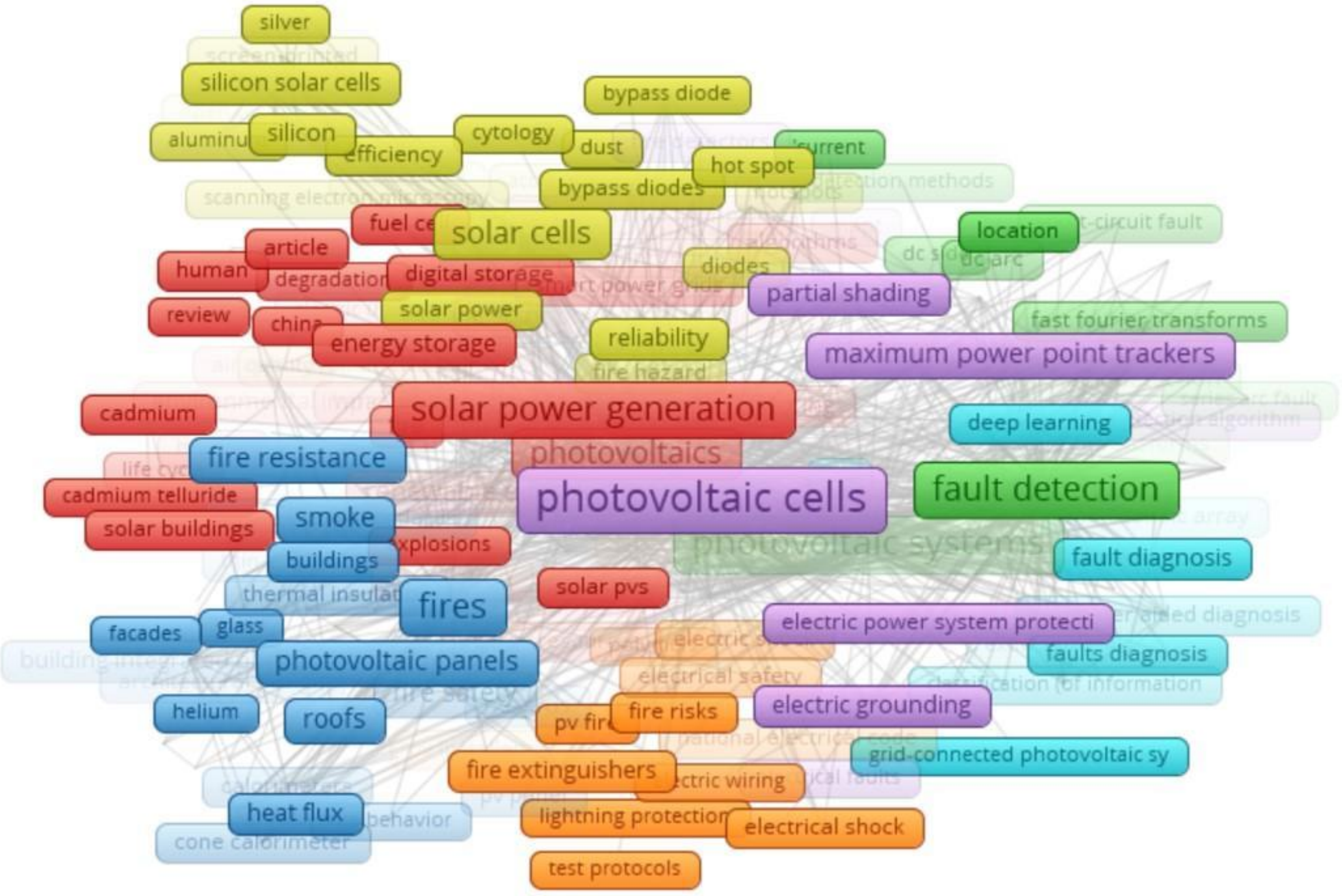

**Figure 8.** Research clusters of Fire and Photovoltaic.

**Table 3.** The main keywords used by the communities detected in the topic Fire and Photovoltaic.

| Cluster | Colour | Keywords by cluster | Main Keywords | Topic |
|---|---|---|---|---|
| **1** | Red | 116 | Solar power generation, monitoring, photovoltaics, installation, cadmium, cadmium telluride, China, energy storage, fire, renewable energies, environmental impact, heating, solar buildings | Fire and Energy Storage |
| **2** | Green | 46 | Photovoltaic systems, fault detection, DC arc, DC side, short-circuit fault, detection methods, detection methods, series arc fault, fast Fourier transform, microgrids | PV faults |
| **3** | Blue | 46 | Photovoltaic panels, fire resistance, buildings, thermal insulation, building integrated photovoltaics, smoke, PV panels, hazards, ventilation, smoke spreads, calorimeters, heat flux | Fire resistance |
| **4** | Yellow | 42 | Fire hazard, reliability, diodes, hot spot, bypass diodes, dust, silicon solar cells, silver, aluminium, efficiency, open circuit voltage, heat transfer, silicon solar cells | Fire hazard |
| **5** | Purple | 38 | Photovoltaic cells, fire hazards, electrical faults, electric grounding, maximum power point trackers, MPPT, partial shading, fire detectors, electric inverters | Fire detectors |
| **6** | Cian | 27 | Fault diagnosis, grid-connected photovoltaic system, computer aided diagnosis, deep learning, learning systems, artificial neural network, internet of things, photovoltaic array | Deep learning |
| **7** | Orange | 21 | Fire risk, PV fire, lighting protection, fire extinguishers, electrical safety, electric shock, polymers, electric wiring, pv array, test protocols | Fire safety |

### 5.1. Fire and energy storage.

Cluster 1 is focused on fire and energy storage, see figure 9. Fire safety is paramount in the context of energy storage systems, where the integration of advanced technologies presents both opportunities and challenges. As the demand for efficient energy storage solutions grows, ensuring robust fire prevention measures becomes crucial to mitigate potential risks and enhance the overall safety of these innovative systems. Balancing the pursuit of sustainable energy storage with rigorous fire safety protocols is essential for a resilient and secure energy landscape. Energy storage systems have evolved significantly in recent years, with a variety of emerging technologies such as lithium-ion batteries, thermal storage, hydroelectric storage and others (Coban et al., 2022). Each of these systems has its own characteristics in terms of storage capacity, efficiency and cost. However, regardless of the type of technology used, fire safety remains a key concern. Recent advances in research into the fire safety of energy storage systems have led to the development of more sophisticated techniques to detect and mitigate the risk of fire. For example, continuous monitoring systems have been implemented that can detect anomalies in temperature or gas release that could indicate a potential safety problem. In addition, safer encapsulation materials and designs are being researched to protect batteries and other system components from thermal failure or short circuits. The integration of advanced energy management technologies is also helping to improve fire safety in energy storage systems. These systems can actively monitor and control the flow of energy, optimising operations to minimise the risk of overcharging or overheating that could trigger a fire. In this line, Li et al. (2021) discuss the broad application potential of Li-ion battery (LIB) energy storage technology due to its longevity, reliability, and adaptability. However, safety concerns impede its rapid industry expansion. The article examines LIB fire characteristics, thermal runaway mechanisms, causes, and monitoring methods, along with current fire protection technologies and safety standards. It concludes with recommendations for developing fire-fighting technology for containerized LIB energy storage systems, serving as a valuable reference for future research in this field.

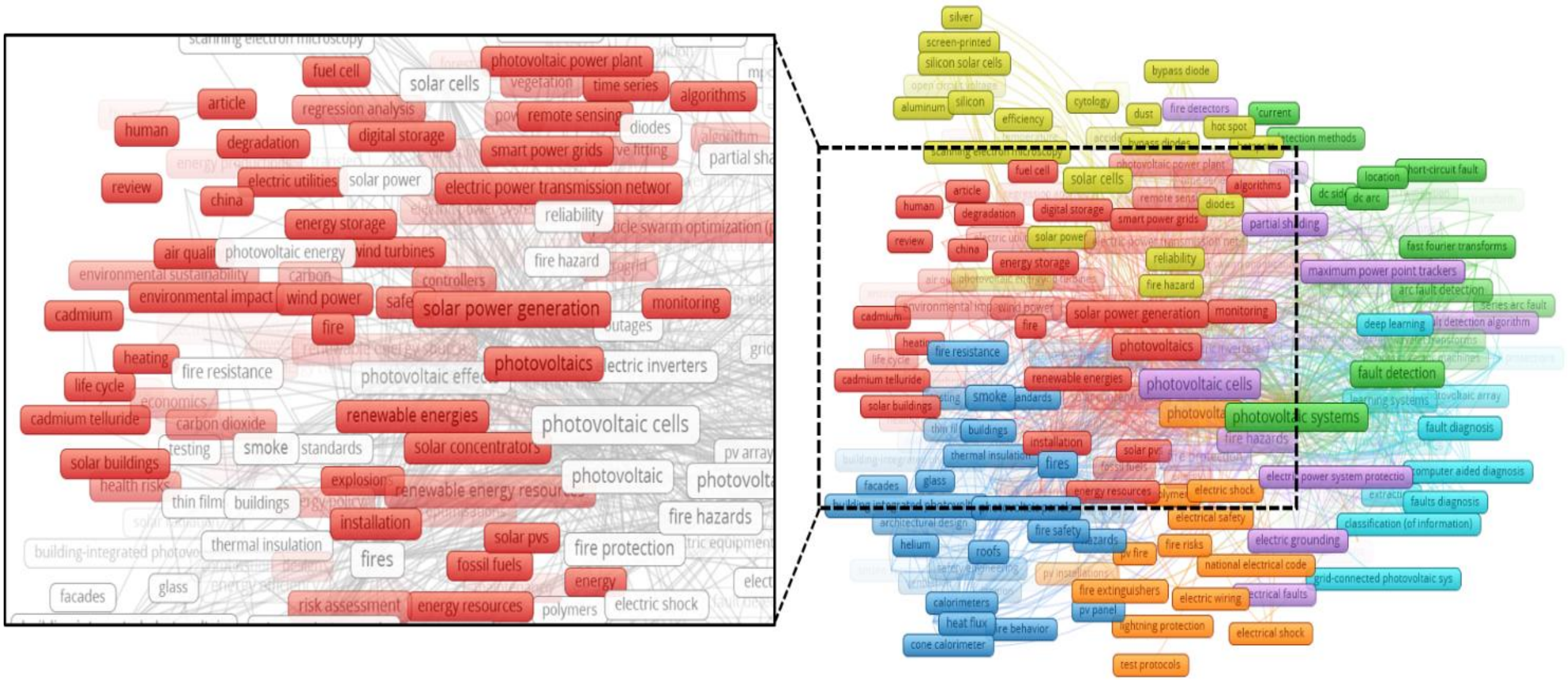

**Figure 9.** Fire and energy storage (Cluster 1).

The most cited study focuses on the importance of improving the detection and mitigation of faults in photovoltaic (PV) systems, highlighting the need to enhance current methods to prevent such accidents. The review identifies limitations in fault detection and mitigation and presents commercially available modern technologies (Alam et al., 2015). Among the available techniques are machine learning approaches to detect faults, using measurements such as voltage, current, and irradiance. However, the challenges of supervised learning models are highlighted due to the requirement for extensive labelled data. To address the issues presented by these techniques, a graph-based semi-supervised model is proposed for detecting and classifying faults with limited labelled data, allowing for simpler and autonomous updates over time, demonstrating effectiveness in fault detection and classification (Zhao et al., 2014). In this context, the use of black phosphorus is proposed as an alternative for manufacturing flame retardants based on BP nanohybrids and multi-layer carbon nanotubes. These nanocomposites showed a significant reduction in heat and flammable volatile release, as well as an improvement in fire safety and environmental stability compared to others (Zou et al., 2020).

### 5.2. Failures in photovoltaic systems

Cluster 2 is focused on failures in photovoltaic systems, see figure 10. Understanding failures in photovoltaic systems is critical for optimizing performance and ensuring the reliability of solar energy. Investigating the causes and mechanisms of failures allows for the development of preventive measures and innovative solutions to enhance the overall resilience of photovoltaic installations. Ongoing research in this field contributes to a more sustainable and efficient utilization of solar power in the global energy landscape. The rapid detection of failures in photovoltaic system components is essential due to its impact on efficiency, energy production, among other reasons. Research in this cluster includes studies on the degradation of solar cells, failure analysis of PV inverters, evaluation of the resistance of PV modules to extreme weather conditions, and development of advanced monitoring techniques to detect early failures in PV systems. The aim of this research is to improve the reliability and durability of PV systems, contributing to greater confidence in solar energy as a clean and sustainable energy source. In this sense, Wang et al. (2020) highlight the threat of DC arc faults to PV and energy storage systems' safety. Existing fault detection methods rely on passive approaches, vulnerable to external factors. They propose an active detection method analyzing current signal responses on the DC bus under high-frequency signal injection. Simulation models validate its effectiveness and employ wavelet transform analysis for fault detection.

The most cited article in cluster 2 analyses the advantages and limitations of fault detection and diagnosis methods in photovoltaic arrays, focusing on accuracy in detecting, locating, and classifying faults (Mellit et al., 2018). In this context, the authors of the most cited study present a maximum power point tracking (MPPT) method for photovoltaic systems (PV) under partial shading conditions using the firefly algorithm, which stands out for its computational simplicity, fast convergence, and execution on an economical microcontroller. The authors highlight the algorithm's advantage over traditional methods like P&O and PSO (Sundareswaran et al., 2014). Other authors focus their study on protection against arc-fault in photovoltaic systems connected in parallel. In the study, they propose three alternative methods to differentiate between series and parallel arc-faults, along with suggestions for mitigation (Johnson et al., 2012).

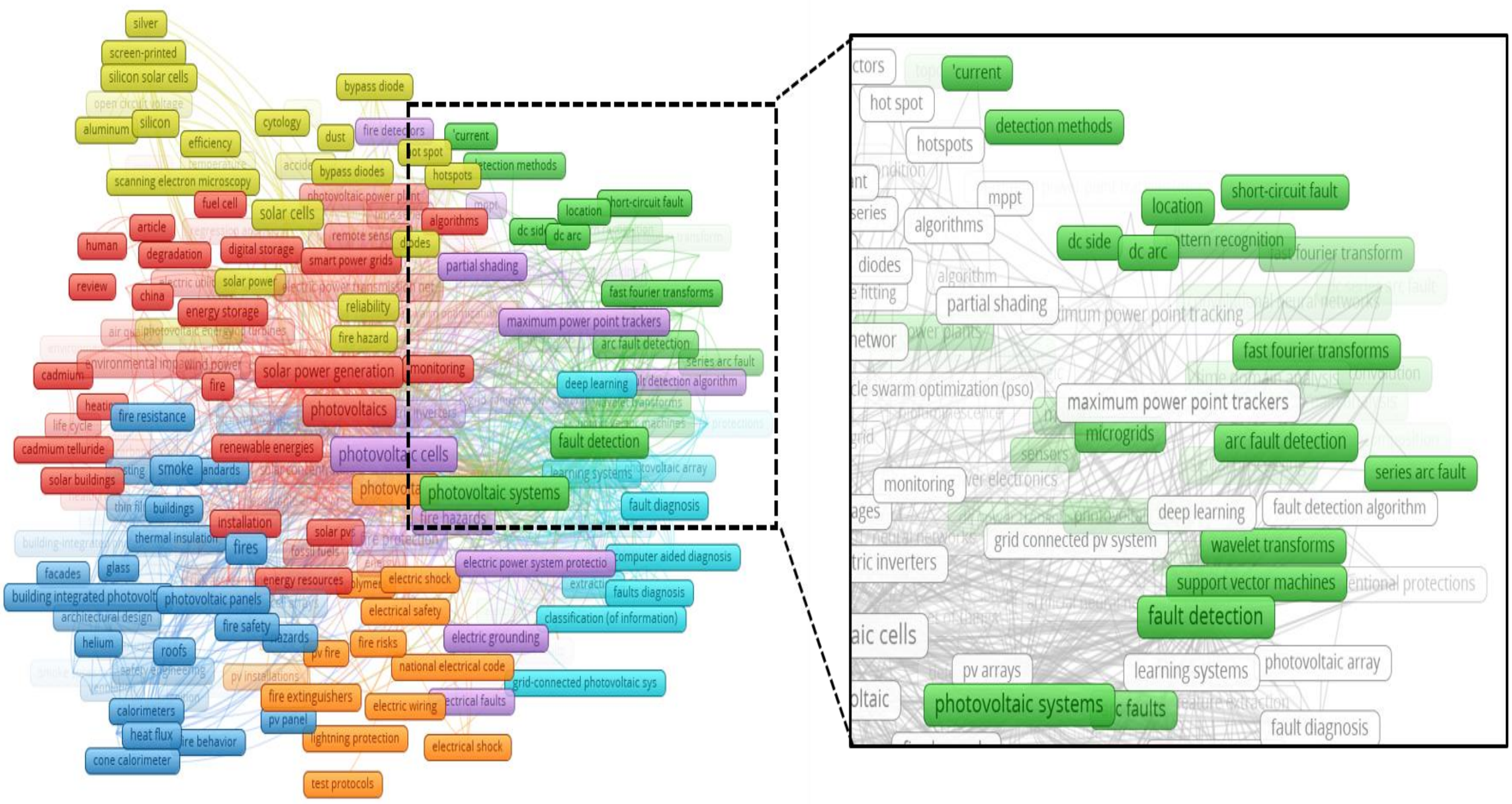

**Figure 10.** Failures in photovoltaic systems (Cluster 2).

### 5.3. Fire resistance

Cluster 3 focuses on fire resistance, see figure 11. Addressing on-fire resistance is paramount in various industries, where the ability of materials and structures to withstand and mitigate fire incidents is crucial for safety and operational continuity. Research and advancements in on-fire resistance technologies contribute to creating more robust and secure environments, ensuring the protection of assets and lives. Developing effective strategies for on-fire resistance is central to fostering resilience and sustainability across diverse sectors. Hot spots are a problem in photovoltaic panels when, among other causes, there is partial shading due to dust or shaded areas. This causes an inadequate distribution of lighting on the panel. The most referenced paper in cluster 3 study the impact of partial and complete shading on poly and monocrystalline photovoltaic modules, with and without the presence of panel bypass diodes installed in hydrocarbon fields where public grid energy is unavailable (Pandian et al., 2016). Wohlgemuth and Kurtz (2012) focus on the fire risk of panels in residential and commercial buildings, suggesting redundant connection designs, robust mounting methods, and changes in safety standards to produce safer modules. The research conducted by Yang et al. (2021) emphasises the significance of investigating fire resistance in photovoltaic modules. The findings indicate that PV modules can be flammable and pose fire risks, highlighting the need to understand and address these risks to ensure safety in solar energy usage. This study underscores the importance of implementing appropriate safety measures and provides a basis for future research aimed at enhancing the fire resistance of photovoltaic systems.

Another significant aspect of this cluster is the scientific progress in reducing fire hazards linked to residential rooftop PV installations. This involves researching advanced materials for solar panels with improved fire-resistant properties, as well as optimizing energy management systems and quick disconnection devices to minimize the spread of fires (Murata et al., 2003). Huang et al. (2018) tested the fire resistance of BIPV (Building integrated photovoltaic) modules according to ISO 834-1:1999, CNS 14803-2010 and ISO 3008:2007 to provide the standard heating curve, in order to investigate the fire resistance characteristics based on the bursting behaviour and surface temperature distribution of BIPV modules type I and type II.

Automated fire detection and extinguishing systems specifically designed for PV environments are being explored to enable rapid and effective emergency response (Verma et al., 2023). It is crucial to understand the propagation of smoke from fires in photovoltaic installations. Studying how smoke disperses in such environments can offer insights into fire behavior in buildings with solar panels. This can aid in developing more effective fire detection and suppression strategies that are tailored to the unique challenges posed by PV systems. By examining smoke propagation, researchers can identify potential hazards, improve evacuation procedures, and enhance overall safety measures to protect both property and lives. In this line, Zhang et al. (2022) conducted computational fluid dynamics (CFD) simulations to analyze smoke spread in solar roof fires, addressing a gap in existing research. The study revealed that roof slope significantly influences smoke propagation, with steeper slopes leading to slower smoke infiltration into buildings. This insight emphasises the significance of considering both fire protection and PV energy performance in solar roof design, especially in low latitude regions where optimal PV angles may increase the risk of smoke infiltration. In other important study, Despinasse and Simone (2015) proposed a new method for testing the fire performance of photovoltaic (PV) modules on roofs. The aim was to simplify existing standards. The study replaced traditional wooden burning brands with a propane burner, which resulted in similar fire behaviour. The authors evaluated smoke evolution, burning drips, flaming debris, and burn-through time for various PV panel types and burner configurations. The results indicate that the dependency on burner output, panel construction, and fire source position should be taken into account when improving fire safety standards in PV module testing. Recent research shows that the greatest risk of smoke spread in PV roof fires is at 15 degrees of roof angles, with 45 and 60 degrees of roof angles being the safest (Zhang et al., 2023).

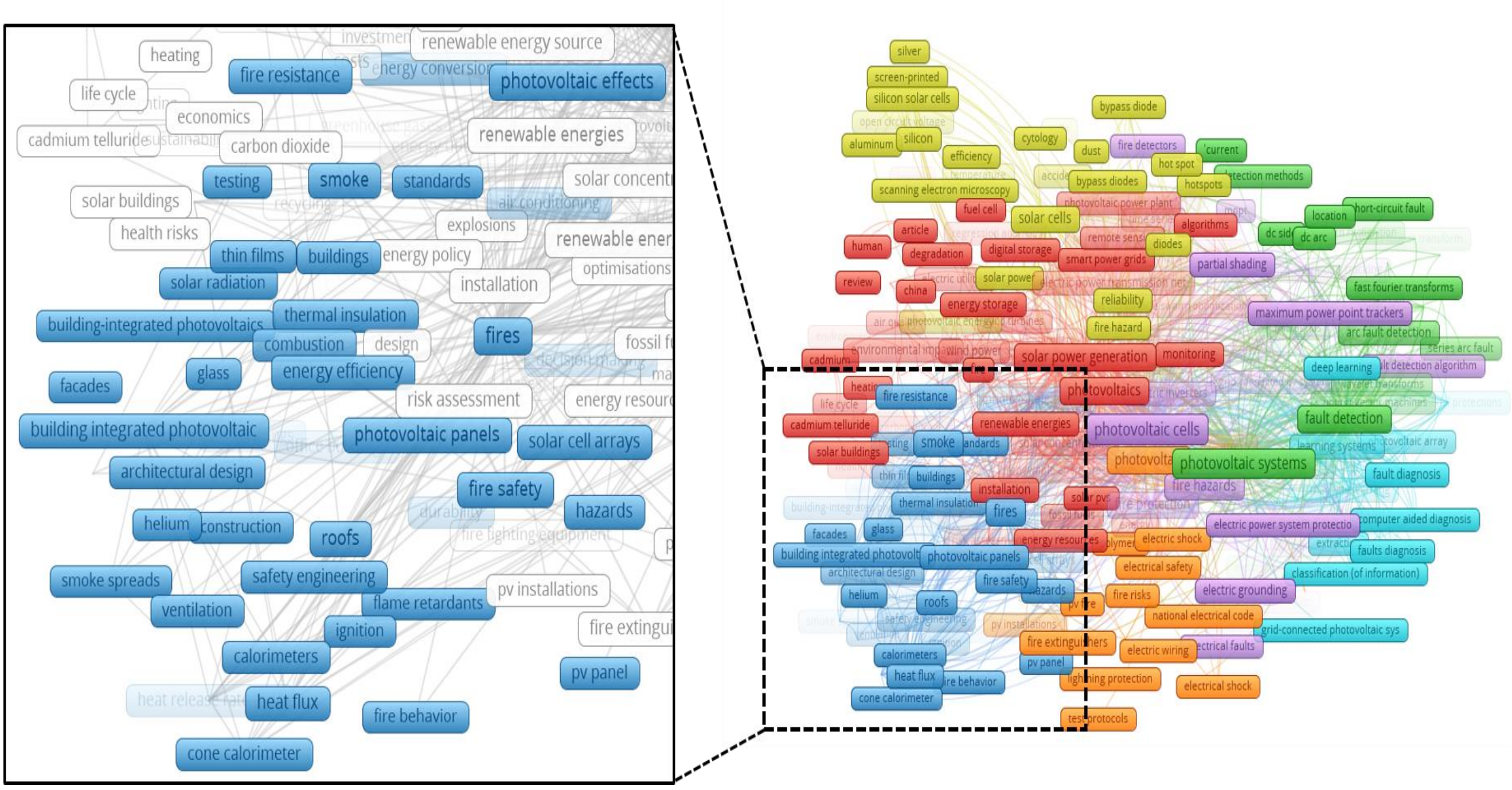

**Figure 11.** Failures in photovoltaic systems (Cluster 3).

### 5.4. Fire hazard

Cluster 4 focuses on fire hazard, see figure 12. Understanding and mitigating on-fire hazards is essential for safeguarding lives, property, and the environment. Research and innovations in fire hazard assessment contribute to the development of effective preventive measures and emergency response strategies. Addressing on-fire hazards is a critical aspect of creating safer and more resilient spaces across various industries and communities. The authors of the most cited article conduct a systematic review of DC arc fault detection techniques in photovoltaic systems, comparing different methods in detail (Lu et al., 2018). Another risk in photovoltaic modules is hot spots, and Guerreiro et al. (2019) present a suitable bypass circuit to completely avoid the occurrence of this problem. Focusing on the same issue, Chen et al. (2019) propose a linear space-time projection method, preserving local geometric structure concerning time series.

Other renewable energy sources can also present fire hazards. In the case of wind energy, the maintenance activities themselves are the main potential source of fire risk, which can lead to the generation of fire outbreaks. Since 2000 to 2014, 235 wind turbine fires have been documented according to CWIF (Caithness Windfarm Information Forum) data. These fires can cause serious damage to the surrounding natural environment. Additionally, due to its height, this type of fire cannot be extinguished by firefighters. During storm conditions, burning debris can spread over a wide area, and the risk of fire is higher in forested areas, and even higher when the ground and vegetation are dry, as in areas with very hot summers. E.g. In July 2019, a turbine fire caused the Juniper Fire (USA), which burned more than 100 hectares over three days. In the same year, a turbine also caused the Rhodes Ranch 3 fire in Abilene, Texas, which burned more than 80 hectares. A fire can cause a significant amount of damage to a wind turbine, resulting in a total loss of the turbine. The cost of repairing or replacing the turbine can range from $1 million to $5 million for onshore models and $6 million to $8 million for offshore turbines.

Of all renewable energy sources, biomass has the highest fire risk. Biomass fuel is consumed in a fire, and biomass storage facilities can be destroyed, often resulting in a long period of production interruption. In addition, the high smoke emissions from biomass fires can harm people and the environment. The risk of self-ignition of stored biomass is essentially a function of the moisture present (Krigstin & Wetzel 2016) and the development of micro-organisms that cause the temperature of the pile to rise (Ferrero et al., 2008). The use of fire detectors and strict control of storage conditions can limit this risk (Sheng & Yao2022).

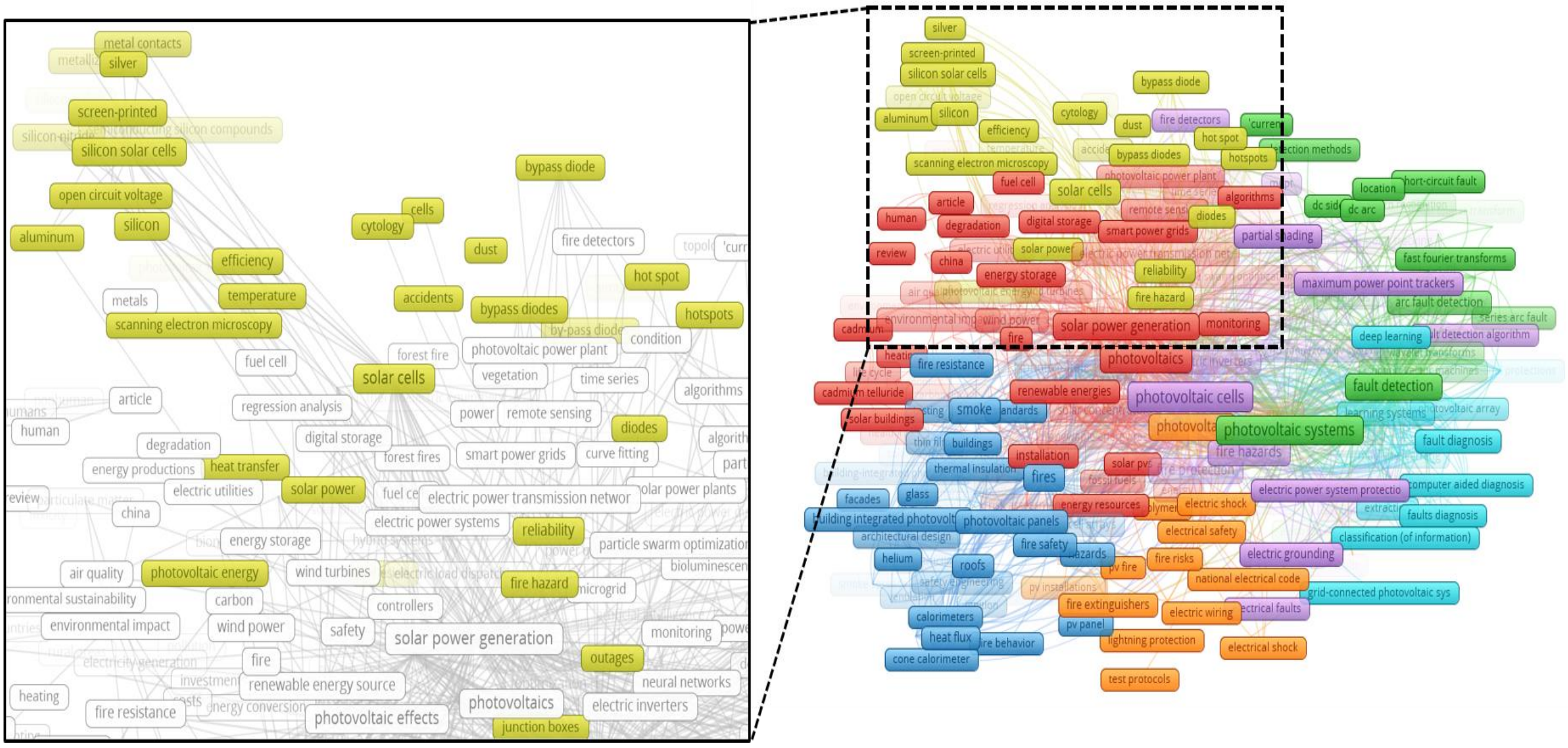

**Figure 12.** Fire hazard (Cluster 4).

### 5.5. Fire detectors

Cluster 5 is centered on fire detectors, see figure 13. The development of advanced on-fire detectors plays a pivotal role in ensuring swift and accurate fire detection, enhancing safety across diverse environments. Continuous innovation in fire detection technologies contributes to early warning systems, minimizing damage and facilitating prompt response measures. The evolution of on-fire detectors is instrumental in creating more secure and resilient spaces, mitigating the impact of fire incidents in various industries and settings. Advances in fire detectors in photovoltaic systems have been significant in recent years. The implementation of AI algorithms has improved the accuracy and speed of fire detection in PV systems. AI can analyse data in real time to identify patterns and anomalies associated with fires, facilitating a rapid response (Akran et al., 2022). More sensitive and specific sensors have also been developed to detect changes in temperature, humidity and other environmental parameters that may indicate the presence of a fire in a PV system. These sensors can provide early warnings, enabling preventive action. Noteworthy is the study carried out by Bagherzadeh et al. (2022), who propose a protection method for detecting electric faults in PV arrays, mitigating quality degradation, power loss, and fire hazards. The method involves detecting faults using current variations, discriminating them from other disturbances based on PV cell transient responses, and locating faulty strings with minimal sensors. Simulation results verify high speed, accuracy, and affordability across various conditions.

Fault detection and diagnosis (FDD) methods are essential for operation and safety in photovoltaic plants. Mellit et al. (2018) review and discuss FDD methods, showing advantages and limitations regarding viability, complexity, cost-effectiveness, and their ability to be implemented on a large scale. Yi and Etemadi (2017) propose a fault detection algorithm based on multiple signal decomposition (MSD) using only voltage and current data from the photovoltaic panel set to detect line-to-line faults. Other renewable energy sources also use fire alarm systems, as example for offshore wind, where the primary means of alarm acknowledgement is the inspiratory smoke detector and the point-type composite smoke and temperature detector farms (Yang et al., 2020). In biomass, for example, the measurement of gas emissions from incipient self-ignition of solid biofuels can be used as an early detection method (Fernandez-Anez et al., 2021).

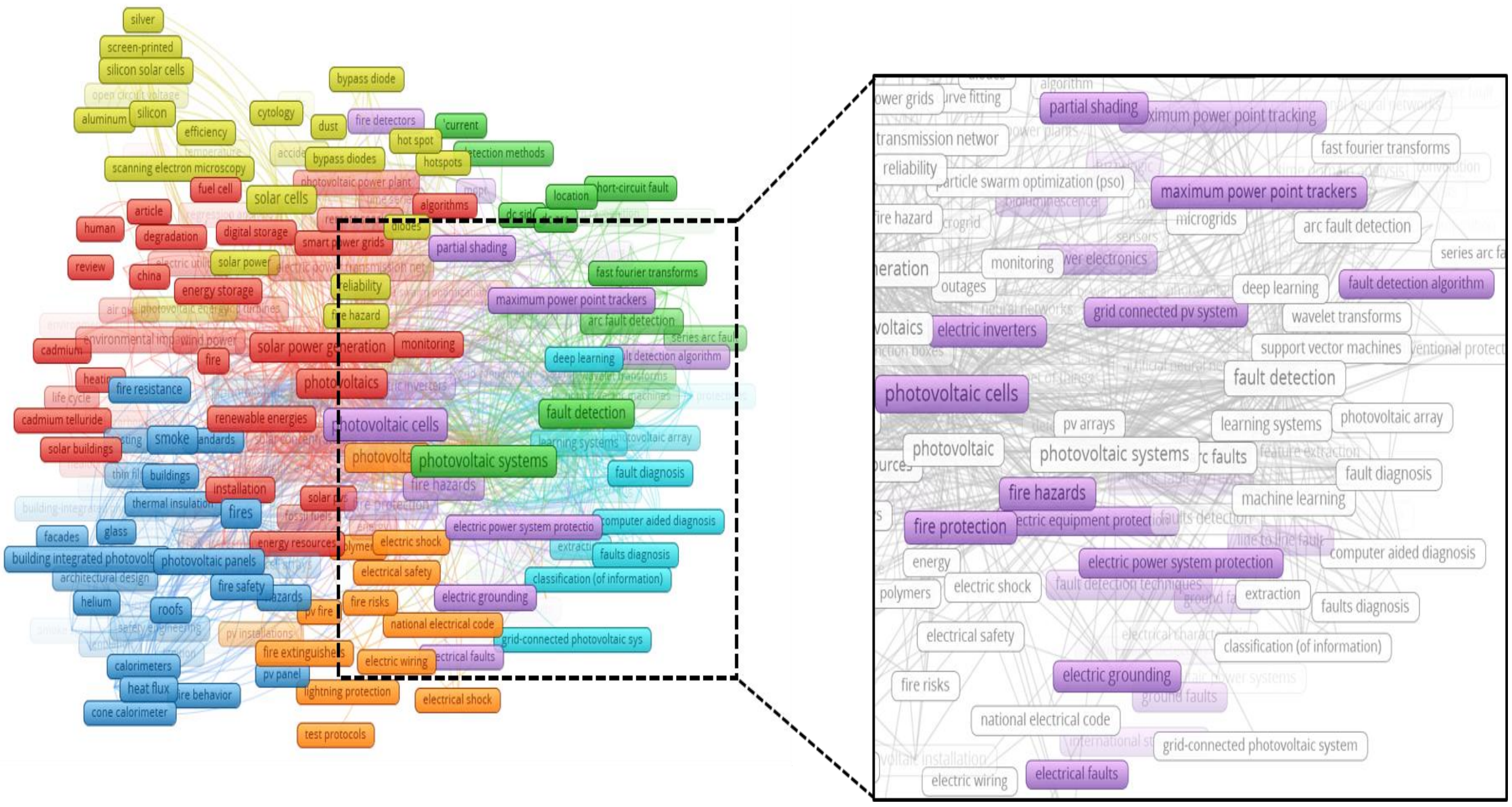

**Figure 13.** Fire detectors (Cluster 5).

### 5.6. Deep Learning

Cluster 6 focuses on Deep Learning for detecting faults in photovoltaic panel installations, see figure 14. The application of Deep Learning in detecting faults in photovoltaic panel installations represents a cutting-edge approach with transformative potential. Leveraging neural networks and advanced algorithms, Deep Learning enables the automated identification and diagnosis of issues, enhancing the efficiency and reliability of solar energy systems. Harnessing the power of artificial intelligence for fault detection not only improves maintenance processes but also contributes to the optimization of overall photovoltaic performance. Pillai et al. (2019) study and classify advanced fault detection methods. The classification of methods is done considering detected faults, fault detection time, among other factors. Lu et al. (2019) propose a convolutional neural network model for fault detection and classification in panels. In this regard, but employing long short-term memory networks, is the proposal of Appiah et al. (2019) for fault detection.

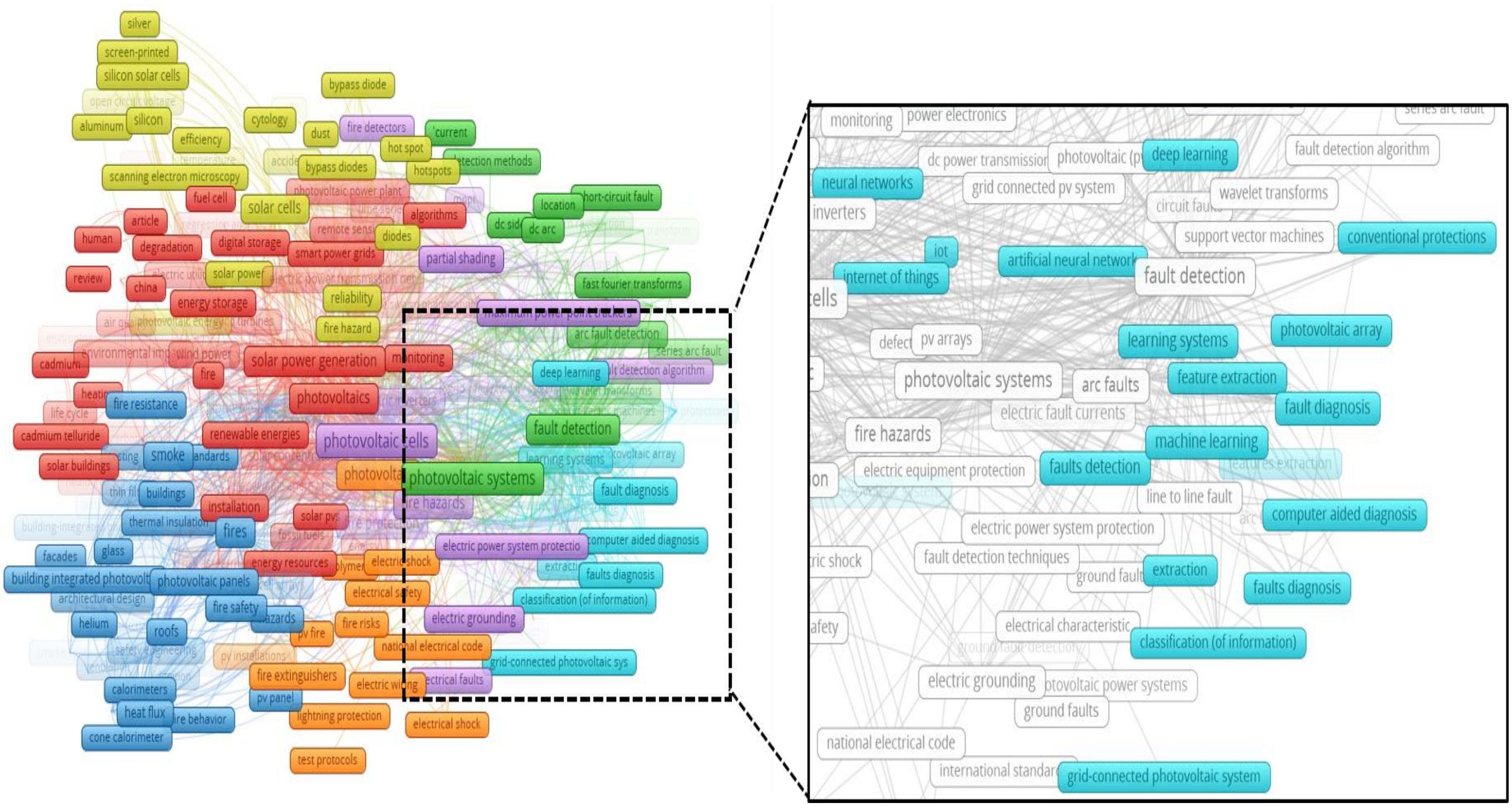

**Figure 14.** Deep Learning (Cluster 6).

### 5.7. Fire safety

The final cluster group address fire safety measures, see figure 15. Recognizing the fire risks linked with photovoltaic energy is crucial to guaranteeing the safety and dependability of solar power systems. With the increasing use of photovoltaic technology, pinpointing and mitigating fire risks becomes paramount to safeguarding installations and averting potential damage. Thoroughly examining these risks is vital for formulating strong safety procedures and promoting the safe integration of photovoltaic energy into our sustainable energy framework. Akram et al. (2022) comprehensively discuss different types of failures occurring in photovoltaic modules and their detection methods in terms of their application and suitability for specific problems. The fire risks associated with photovoltaic modules and their mitigation are also addressed. Additionally, the combined application of these methods is reviewed in detail. Reil et al. (2012) proposed different tests to classify risks in cell connectors that transform into arc risk in photovoltaic panels, serving as a quality control. Aram et al, (2021) review the state of the art of fire safety of PV systems in buildings. Their study concludes that a PV fire incident is a complex and multifaceted phenomenon that cannot be reduced to a single variable that causes a single outcome. It is noticeable that there is no reflection of roof fires, although there are some works which do reflect their importance (Kristensen, & Jomaas, 2018) in our analysis, they do not appear as a cluster, so this should be a line of work for consideration; where works have been done in this area in the recent past, especially related to fire safety (Juarez-Lopez et al., 2023).

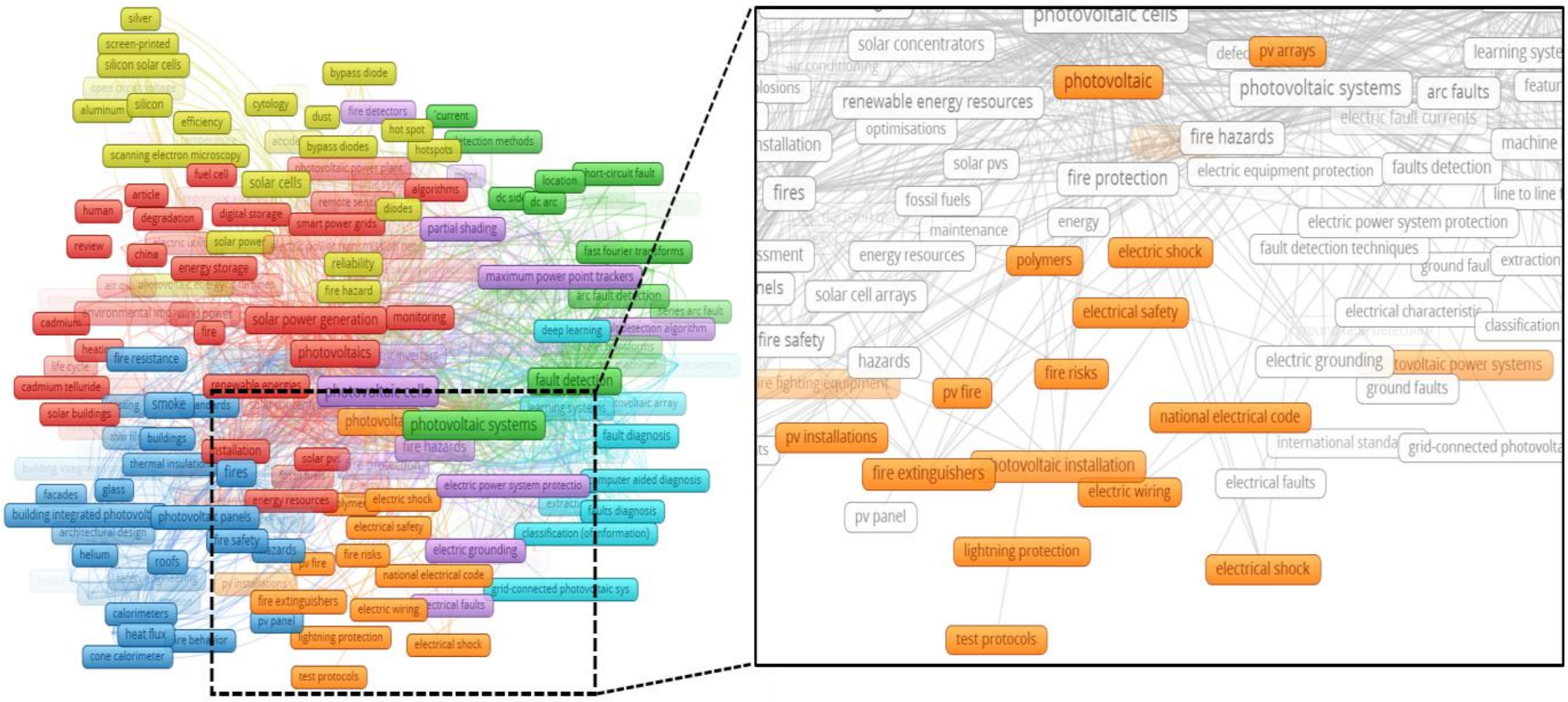

**Figure 15.** Fire safety (Cluster 7).

## 6. Future research directions

The literature has shown that the fire risk and associated hazards in photovoltaic installations are lower than those of other renewable energies, such as biomass or wind energy. Identifying fires caused by fault in photovoltaic installations in the literature can be challenging. This lack of detail makes it difficult to study these fires and determine how to address them and should therefore be considered as a line of future research. Therefore, it is recommended to identify them in the causes of the fires.

It is recommended that regulations and safety standards be created to mitigate the risk of fire and ensure the safety of both installers and end-users during the installation and operation of PV systems. The review conducted confirmed that there is a shortage of literature on PV roofs and fire. In view of the potential for expansion in this area, it is necessary to address the issue of specific binding regulations in relation to fire resistance for PV roofs. Furthermore, it is necessary to conduct research and develop safer and more fire-resistant materials for key components of PV systems, such as solar modules, cables and inverters.

Another area with potential fire risk is agrivoltaics, where photovoltaic installations will be located in natural environments where firefighting can be difficult. Consequently, this opens up new perspectives for research into fires in photovoltaic installations and thus helps to contribute to a more sustainable future through the use of clean energy technologies such as photovoltaics. This will involve investigating how fire spreads through solar panels, mounting structures and electrical components, and developing more effective prevention and mitigation measures based on this knowledge. In addition, the effect of vegetation on the flammability of agrivoltaic systems should be studied, considering how surrounding vegetation can act as an additional fuel or barrier to the spread of fire. Another important approach is to develop fire-resistant materials specifically for PV system construction to reduce the risk of ignition and minimise damage in the event of a fire. Finally, it is very important to explore the integration of fire detection and suppression technologies, such as remote monitoring systems and automatic extinguishing systems, to improve emergency response.

**7. Conclusions**

In this study, an analysis of all scientific publications related to fire research in photovoltaic systems that were published between 1983 and 2023 has been carried out. The analysis focuses specifically on PV-related fires and reveals a significant growth in research output. The early years indicated an incipient stage, with a notable increase from 2011 onwards, peaking between 2018 and 2022, a growth that is parallel to the installation of photovoltaics worldwide. The engineering scientific category accounts for the largest share of publications on this topic (34%), followed by energy (17%) and materials science and computer science (both 10%). Global research output indicates that the United States is the leading country, followed by China and India, each with different research priorities. The topic is being researched by three main institutions: Sandia National Laboratories in New Mexico, the University of Science and Technology of China, and the National Renewable Energy Laboratory in Colorado.

The scientific landscape of photovoltaic (PV) fire research has evolved over the years. Initially, from 2013 to 2015, the focus was on practical applications of PV technology, safety, materials, and system reliability. From 2016 to 2017, the focus shifted towards examining specific fire concerns, such as the impact of PV panels on roofs and material risks. Between 2018 and 2019, the focus shifted towards technical aspects such as fault detection, grid integration, and vulnerabilities on the DC side of the systems. From 2020 to 2023, progress was made towards sophisticated diagnostics, including fault diagnosis, maximum power point tracking (MPPT), and deep learning techniques to improve safety and efficiency. This progression suggests a maturing understanding of PV fire risks, moving from general safety considerations to addressing specific technical challenges, in parallel with the wider integration of PV technology across different sectors.

In this bibliometric study, the scientific communities around which the publications are grouped are identified on the basis of the keywords of the articles.

The analysis of these clusters of scientific communities makes it possible to identify dominant topics, emerging areas of interest and the trajectory of research development over time. Seven scientific communities have been detected in which these works are grouped according to their keywords: Fire and Energy Storage, PV faults, Fire resistance, Fire hazard, Fire detectors, Deep learning, and Fire safety.

Fire and Energy Storage Cluster emphasises robust prevention measures in energy storage, proposing innovations in machine learning and flame retardants. PV faults cluster highlights the importance of fault prevention, introducing fault detection and differentiation methods. Fire resistance cluster elaborates strategies on fire resistance, addressing hot spots and suggesting design modifications for safer PV modules. Fire hazard cluster analyses fire hazards, evaluating detection techniques and proposing innovative solutions. Fire detectors cluster focuses on advanced fire detectors, highlighting fault detection and diagnostic methods for early warning systems. Deep learning cluster pioneers the use of Deep Learning for fault detection, employing neural network models for automated problem identification. Finally, Fire safety cluster explores the fire risks associated with photovoltaic energy is essential to ensure the safety and reliability of solar power systems.

Acknowledgments: The authors would also like to thank the University of Córdoba (Spain) through PPIT_2022E_026695 “Nuevas Soluciones para la Sostenibilidad Energética y Ambiental en Edificios Públicos”.